\pdfoutput=1

\documentclass[10pt]{article}

\usepackage{listings}    % for snippets of code using \begin{lstlisting}. use "escapechar=©" when needing \label
\usepackage{graphicx}    % for figures, eg, \includegraphics
\usepackage{subcaption}  % for subfigures, ie figures with internal figures
\usepackage{algorithm}   % for displaying algorithm in pseudo-code, using \begin{algorithm}
\usepackage{algorithmic} % and \begin{algorithmic}
\usepackage{authblk}     % for \affil
\usepackage{booktabs}    % for better tables, eg using \toprule
\usepackage{hyperref}    % to handle URL links with \url
\usepackage[T1]{fontenc} % needed for scaling fonts
\usepackage{lmodern}     % same as above, for scalable fonts
\usepackage[htt]{hyphenat} % for line breaks in texttt
\usepackage[square,numbers]{natbib} % needed for the references, when using ACM-Reference-Format outside an
\usepackage{microtype} % ligatures are extremely annoying, especially when copy&paste text from PDF
\DisableLigatures{}    % https://tex.stackexchange.com/questions/439651/how-do-i-disable-ligatures

\usepackage{geometry}
\usepackage{xcolor}
\definecolor{codegreen}{rgb}{0.25,0.5,0.35}
\definecolor{codegray}{rgb}{0.5,0.5,0.5}
\definecolor{codepurple}{rgb}{0.6,0,0}
\definecolor{backcolour}{rgb}{0.95,0.95,0.92}
\definecolor{colorstring}{rgb}{0.5,0,0.35}
\definecolor{rltred}{rgb}{0.5,0,0}
\definecolor{rltgreen}{rgb}{0,0.5,0}
\definecolor{rltblue}{rgb}{0,0,0.5}
\definecolor{DarkGreen}{rgb}{0.00,0.60,0.00}
\definecolor{ScarletRed}{rgb}{0.80,0.00,0.00}
\definecolor{blizzardblue}{rgb}{0.67, 0.9, 0.93}
\definecolor{green-yellow}{rgb}{0.68, 1.0, 0.18}
\definecolor{dkgreen}{rgb}{0,0.6,0}
\definecolor{gray}{rgb}{0.5,0.5,0.5}
\definecolor{mauve}{rgb}{0.58,0,0.82}
\definecolor{lightgrey}{rgb}{0.90,0.90,0.90}
\definecolor{grey}{gray}{0.75}
\definecolor{light-gray}{gray}{0.80}

\lstdefinestyle{mystyle}{
	backgroundcolor=\color{backcolour},
	commentstyle=\color{codegreen},
	keywordstyle=\color{colorstring}\bfseries,
	numberstyle=\ttfamily\color{codegray},
	stringstyle=\color{codepurple},
            basicstyle={\scriptsize\ttfamily},
	breakatwhitespace=false,
	breaklines=true,
	captionpos=b,
	keepspaces=true,
	numbers=none,
	numbers=left,
	numbersep=2pt,
	showspaces=false,
	showstringspaces=false,
	showtabs=false,
	tabsize=2
}
\usepackage{xspace}
\newcommand{\evo}{{\sc EvoMaster}\xspace}
\newcommand{\etal}{{\emph{et al.}}\xspace}

\usepackage{boxedminipage}
\newenvironment{result}%
{\smallskip
	\noindent
	\let\emph=\textbf
	\begin{boxedminipage}{\columnwidth}\begin{center}\em}%
		{\end{center}\end{boxedminipage}%
}

\usepackage{ifthen}
\newboolean{showcomments}
\setboolean{showcomments}{true} % comment this line to deactivate comments

\ifthenelse{\boolean{showcomments}}{
	\newcommand{\nbc}[3]{
		{\colorbox{#3}{\bfseries\sffamily\scriptsize\textcolor{white}{#1}}}
		{\textcolor{#3}{\sf\small$\langle$\textit{#2}$\rangle$}}}
	
}{
	\newcommand{\nbc}[3]{}

}

\title{White-Box and Black-Box Fuzzing for GraphQL APIs}  % use \\ to break line, if needed
\author{Asma Belhadi, Man Zhang and Andrea Arcuri \\Kristiania University College, Norway}
\date{}

\begin{document}

\maketitle

\begin{abstract}

The Graph Query Language (GraphQL) is a powerful language for APIs manipulation in web services.
It has been recently introduced as an alternative solution for addressing the limitations of RESTful APIs.
This paper introduces an automated solution for GraphQL APIs testing.
We present a full framework for automated APIs testing, from the schema extraction to test case generation.
In addition, we consider two kinds of testing: white-box and black-box testing.
The white-box testing is performed when the source code of the GraphQL API is available.
Our approach is based on evolutionary search.
Test cases are evolved to intelligently explore the solution space while maximizing code coverage and fault-finding criteria.
The black-box testing does not require access to the source code of the GraphQL API.
It is therefore of more general applicability, albeit it has worse performance.
In this context, we use a random search to generate GraphQL data.
The proposed framework is implemented and integrated into the open-source \evo tool.
With enabled white-box heuristics, i.e., white-box mode,
experiments on 7 open-source GraphQL APIs show statistically significant improvement of the evolutionary approach compared to the baseline random search.
In addition, experiments on 31 online GraphQL
APIs reveal the ability of the black-box mode to detect real faults.

\end{abstract}

{\bf Keywords}: GraphQL, EvoMaster, Evolutionary Algorithms, Automated Testing, Random, SBST, SBSE, fuzzing.

%%%%%%%%%%%%%%%%%%%%%%%%%%%%%%%%%%%%%%%%%%%%%%%%%%%%%%%%%%%%%%%%%%%%%%%%%%%%

\section{Introduction}

Web services are very common in industry, especially in enterprise applications using microservice architectures~\cite{newman2015building}.
They are also becoming more common with the appearing of smart city technologies, where microservices are largely exploited in industrial internet-of-things settings~\cite{cirillo2020smart,cabrera2020towards}. 
The investigation of automating techniques for generating test cases for web service APIs has become a research topic of importance for practitioners~\cite{arcuri2018experience}. 

Due to the high number of possible configurations for the test cases, evolutionary techniques have been successfully used to address different software testing problems~\cite{ali2009systematic,harman2012search}. 
Common examples are EvoSuite for unit test generation for Java programs~\cite{fraser2011evosuite}, Sapienz for mobile testing~\cite{mao2016sapienz} and \evo for REST API testing~\cite{arcuri2018evomaster}.

The Graph Query Language (GraphQL) is a powerful language of web-based data access, created in $2012$ and open sourced by Meta/Facebook in $2015$~\cite{GraphQLFoundation}. 
It addresses some of the RESTful API limitations, like the possibility of specifying what to fetch on a graph of interconnected data with a single query~\cite{hartig2018semantics,taelman2018graphql}. 
Different companies have started to provide web APIs using GraphQL\footnote{https://graphql.org/users/}, like for example Facebook, GitHub, Atlassian, and Coursera.

To the best of our knowledge, only three recent approaches in the scientific literature exist which explore the automated testing for GraphQL APIs, dealing with ``deviation testing''~\cite{vargas2018deviation}, ``property-based testing''~\cite{karlsson2020automatic} and ``harvesting production queries''~\cite{zetterlund2022harvesting}.
To deal with this gap in the scientific literature, this paper presents the first white-box and black-box test generation approach for automating GraphQL APIs testing.
We developed a full framework for automate GraphQL APIs testing, from the schema extraction to the generation of the test cases in executable test suite files (e.g., using JUnit and Jest). 
The proposed framework has been implemented as an extension to the open-source \evo tool~\cite{arcuri2018evomaster,arcuri2021evomaster}, and it is freely available online.

The main contributions of this research work are as follows:
\begin{enumerate}
    \item We investigate two kinds of testing: white-box and black-box testing. 
    The white-box testing is performed when the source code of the GraphQL API is provided, whereas the black-box testing is employed when the source code of the GraphQL APIs is either missing or when users do not have access to it (e.g., remote services). 
    
    \item  We develop an evolutionary-based search for white-box testing. 
    It intelligently explores the test case space in GraphQL APIs. 
    Different types of genes are created which allow the complete data representation of the GraphQL schema. 
    The genetic operators are used to explore the test case space while maximizing metrics such as code coverage and the number of detected faults. 
    We also adopt a random search for black-box testing, where test cases are generated randomly from GraphQL schema without considering information from the source code of the GraphQL APIs. 
    
    \item To validate the applicability of our presented framework, an empirical study has been carried out on 7 GraphQL APIs with source code, and 31 online  GraphQL APIs. 
    For white-box testing, the results show a clear improvement of using the evolutionary algorithm compared with the random search baseline. 
Line coverage is improved by 6\% on average (up to +26.9\% in one API), and more 
 errors were automatically found in the analyzed APIs (+77 new errors in total).
For black-box testing, the results also reveal the detection of up to 641 endpoints with errors over 825 endpoints. 
\end{enumerate}

This paper is an extension of a short poster~\cite{belhadi2022graphql}.
In~\cite{belhadi2022graphql} we provided an initial implementation to support a subset of GraphQL, with an empirical study of white-box testing on two artificial APIs. 
In this paper, we significantly extended such support (e.g., how to deal with nested function calls), and provided a much larger empirical study, including experiments on black-box testing as well.

The article is structured as follows. 
Section~\ref{sec:background} provides background information on  GraphQL APIs and \evo. 
Section~\ref{sec:relatedwork} discusses related work.
Section~\ref{sec:EvoMaster+} gives a detailed explanation of the main components of the proposed framework. 
The details of our empirical study are presented in Section~\ref{sec:Empirical Study}, followed by a discussion of the main findings of our research work in Section~\ref{sec:Discussion}. 
Section~\ref{sec:threats} discusses the possible threats to validity. 
Finally, Section~\ref{sec:conclusions} concludes the paper.

%%%%%%%%%%%%%%%%%%%%%%%%%%%%%%%%%%%%%%%%%%%%%%%%%%%%%%%%%%%%%%%%%%%%%%%%%%%%%%%%%%%%%%%%%%%%%%%%%%%%%%%%%%%%%%%%%%%%
\section{Background}
\label{sec:background}

This section provides important background information to better understand the rest of the paper, 
in particular regarding GraphQL APIs (Section~\ref{sub:GraphQl}) and the \evo tool (Section~\ref{sec:evomaster}).

%-------------------------------------------------------------------------------------------------------------------
\subsection{GraphQL}
\label{sub:GraphQl}
\begin{figure}
	\centering
	\begin{subfigure}{.4\textwidth}
		\includegraphics[width= \textwidth]{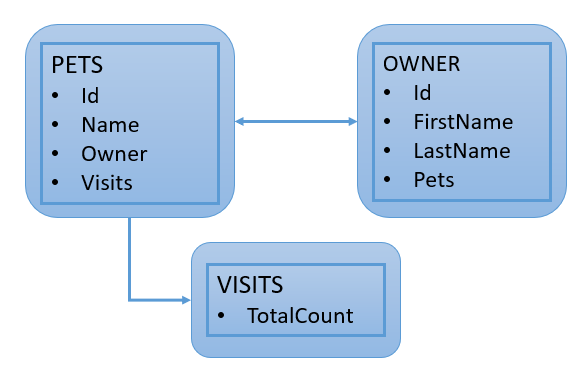}
		\caption{Data graph} \label{fig:DataGraph}
	\end{subfigure}\hspace*{\fill}%
	\begin{subfigure}{.3\textwidth}
		\begin{lstlisting}[language=java]
{
	pets {
		id
		name
		owner {
			id
			firstName
			lastName
			pets {
				id
				name
			}
		}
		visits {
			totalCount
		}
	}
}
		\end{lstlisting}
		\caption{GraphQL query} \label{fig:GraphQLQuery}
	\end{subfigure}
	\caption{Example of GraphQL data graph and a query on it}
\end{figure}

GraphQL is a query language and server-side runtime for application programming interfaces (APIs)~\cite{GraphQLFoundation}.
Given a set of data represented with a graph of connected nodes, GraphQL enables to query such graph, specifying for each node which fields and connections to retrieve (and so recursively on each retrieved connected node).
Figure~\ref{fig:DataGraph} shows a simplified/reduced example of graph for a pet clinic, whereas Figure~\ref{fig:GraphQLQuery} shows a GraphQL query on it to retrieve the list of all pets with their owners and that have been registered in the clinic.

A GraphQL web service would be typically listening on a TCP socket, expecting GraphQL queries as part of  HTTP requests, on a given HTTP endpoint (typically \texttt{/graphql}).
However, GraphQL is not tight to HTTP, and queries could be technically sent via other communication mechanisms.  
One advantage here is that a user can fetch all the data they need (and only that) in a single HTTP call. 

A core concept in GraphQL is the schema, which is a collection of types and relationships between those types.
It describes which kind of types and fields on those types a client can request. 
GraphQL is a strongly typed language and has its own language to write the schema.

Commonly used types definition in a GraphQL schema are as follows:

\begin{enumerate}
\item \textbf{Object types}.
The most frequent elements of a GraphQL schema are object types. 
It indicates which kind of object (e.g., a node in the graph) you can fetch and what fields it has.

\item \textbf{Query type}.
A Query type is a special type in GraphQL that defines entry points (equivalent to remote procedure calls) for fetching data from the graph. 
It is the same as an object type, but its name is always Query. 
Each field of the Query type describes the name and return type of a different entry point.
Note that a GraphQL server has to define a query type. 

\item \textbf{Mutation type}.
A GraphQL operation can either be a read or a write operation. 
A Query type is used to read data while the Mutation type is used to modify data. 
The Mutation type follows the same syntactical structure as queries, however it defines entry points for writing operations. 
Note that a GraphQL server may or may not have a mutation type.

\item \textbf{Scalar types}.
Scalar types are primitive types that resolve to concrete data. 
GraphQL has five default scalar types: Int, Float, String, Boolean and ID. 
Note that GraphQL allows creating custom scalar types for more specific usage.

\item \textbf{Input types}.
Input types are object types that allow passing complex objects as arguments to queries and mutations. 
An input type's definition is alike to an object type's, but it starts with the keyword \texttt{input} instead of type. 
Note that input types can only have basic field types (input types or scalar types) and can not have field arguments.

\item \textbf{Enum types}.
An Enum type is a special scalar type with a restricted set of allowed values specified in the schema.

\item \textbf{Interface types}.
An Interface type is an abstract type. 
An interface is composed of a set of fields held by multiple object types.
When an object type implements an interface, it has to include all of that interface's fields.
Thereby, interfaces enable returning any object type that implements that interface. 
Note that, we can query an interface schema type for any fields defined in the interface itself and we can also query it for fields that are not in the interface but in the object types implementing the interface.

\item \textbf{Union types}.
Like interface types, union type belong to the GraphQL abstract types. 
It allows to define a schema type that belong to multiple types. 
In its definition, a union type will determine which object types are included. 
In this case, the schema field can return any object type that is described by the union.
Note that, all union's included types should be object types (e.g., not Input types).

\item \textbf{Non-Nullable type}.
All types in GraphQL are nullable by default, i.e., the server can return a null value for all the previous types.
To override this default and specify that \texttt{null} is not a valid response, an exclamation mark (!) following the type is added indicating that this field is required.
The Non-Null type can also be used in arguments. 
Note that, in queries a field is always optional, i.e., one can skip a non-nullable field and the query would be still valid. 
However, if a field is required (declared as non-nullable) the server must never return the null value if the query fetches such field. 
With regard to input arguments, by default they are optional. 
However, if a type is declared as non-null, besides  not taking the value null, it also does not accept omission (i.e., the input argument must be present).
\end{enumerate}

The following is a fragment of a GraphQL schema extracted from one of the SUTs used in our empirical study in Section~\ref{sec:Empirical Study} (i.e., \textit{petclinic}~\cite{gqlpetclinic}).
The schema describes the entry point \texttt{pets} that returns a list of all pets that have been registered in the pet clinic.
It is defined as a non-nullable object array type named \texttt{Pet}.
The object type \texttt{Pet} has as a field a non-nullable integer scalar type that represents the \texttt{id} of the pet.
It also defines a non-nullabe object type \texttt{Owner} which implements an interface named \texttt{Person}. 
The field \texttt{VisitConnection} is an object type specifying all the visits to the pet clinic of this pet.
It has an non-nullable integer field \texttt{totalCount} that reports the total number of visits for this pet.  
\begin{lstlisting}[language=java]
type Query {  
    pets: [Pet!]!
}

type Pet {
    id: Int!
    name: String!
    owner: Owner!
    visits: VisitConnection!
}

type Owner implements Person {
    id: Int!
    firstName: String!
    lastName: String!
    pets: [Pet!]!
}

type VisitConnection {
    totalCount: Int!
}

interface Person {
    id: Int!
    firstName: String!
    lastName: String!
}
\end{lstlisting}

As a response to a query, a GraphQL API will return a JSON object with two fields:
\texttt{data} that contains the result of the query, and \texttt{errors} if there was any error with the query (and in this case the \texttt{data} field would not be present).
Notice that GraphQL makes no distinction between user errors (e.g., a wrongly formatted query or an input does not satisfy a business logic constraint) and server errors (e.g., internal crash in the business logic of the API, like a null-pointer exception). 
However, it might support it in the future\footnote{https://github.com/graphql/graphql-spec/issues/698}.
Furthermore, as GraphQL is independent from HTTP, a query with errors could still have a HTTP response 200 (i.e., OK), and this is a common behavior among GraphQL framework implementations. 

Another limitation of GraphQL is that currently it has no standarized way to express constraints on the fields of the graph (e.g., an integer within a specific numeric range, or a string that should satisfy a given regular expression).
Constraints could be added with ``directives'', which are decorators used to extend the semantics of the schema.
But those would be custom, and unique for each different implemented API.

%-------------------------------------------------------------------------------------------------------------------
\subsection{EvoMaster}
\label{sec:evomaster}

\evo~\cite{arcuri2018evomaster,arcuri2021evomaster} is an open-source tool that aims at system test generation, currently targeting REST web services~\cite{arcuri2019restful}.
Internally it uses the evolutionary algorithm (i.e., Many Independent Objective algorithm (MIO)~\cite{mio2017}) enhanced with \emph{Adaptive Hypermutation}~\cite{zhang2021adaptive}, and can handle both black-box~\cite{arcuri2020blackbox} and white-box testing~\cite{arcuri2020testability}.
For white-box testing, it uses established heuristics like the \emph{branch distance}~\cite{Kor90,mio2017,arcuri2019restful}, it employs testability transformations to smooth the search landscape~\cite{arcuri2020testability}, and can also analyze all interactions with SQL databases to improve the fitness function~\cite{arcuri2020sql}.

\evo currently targets REST APIs running on the JVM  and NodeJS~\cite{js2022}(albeit for black-box testing it can be applied on any kind of REST web service), and it outputs test suite files in the JUnit and Jest format.
Each generated test case is composed by one or more HTTP calls, and SQL data to initialize the database (if any).   

When targeting RESTful APIs, \evo analyzes their schema, and create a chromosome representation with a rich gene system to represent all possible needed types (from integers and strings to full JSON objects).
The resulting phenotype will represent complete HTTP requests, where each gene would represent the different decisions that need to be made in these HTTP requests (e.g., query/path parameters in the URLs, and body payloads in POST/PUT/PATCH requests).
\evo evolves test cases based on different metrics, like statement and branch coverage, as well as coverage of the HTTP status codes for each endpoint in the API.
To detect potential faults, it considers the 500 HTTP status code (i.e., server error) and possible mismatches between the schema and the concrete responses~\cite{marculescu2022faults}.

Internally, \evo uses the MIO~\cite{mio2017} algorithm to evolve test cases. 
It is a genetic based evolutionary algorithm, designed specifically for handling system test case generation. 
Here, we briefly discuss how it works, but for the full details of MIO we refer to~\cite{mio2017}.
MIO is a multi-population algorithm, with one population for each testing target. 
MIO evolves individuals that are test cases, and outputs test suites. 
At the beginning of the search, one single population is randomly initialized, based on the chromosome templates constructed from the schema. 
For testing RESTful APIs, several kinds of testing targets are taken into account, like for example statements coverage in the System Under Test (SUT), branch coverage and returned HTTP status codes. 
Next, at each step, the MIO either samples a new test at random or selects one existing test from one population that includes yet to covered targets, and mutate such a test.
Different strategies are used to select which population to sample from.    
Individuals are manipulated through only one operator, which is the mutation operator (i.e., no crossover). 
Two types of mutation are applied: either a structure mutation or internal mutation. 
A test case is composed by one or more ``action'' (i.e., an HTTP call in the case of testing of web services).
In contrast to the internal mutation, which affects only the values of the genes in the actions, for instance flipping the value of a boolean gene from \texttt{True} to \texttt{False}, the structure mutation acts on the structure itself, such as adding and removing actions (e.g., HTTP calls). 
Each time a new test is sampled/mutated, its fitness is calculated. 
If it achieves any improvement on any target (regardless of the population it was sampled from), it will be saved in the corresponding populations (and the worst individuals in such populations are deleted). 
In this context, if a target is covered by a test, it is saved in an archive, the corresponding population is shrunk to one single individual, and it will never expand again nor used for sampling. 
If new targets are reached (but not fully covered) during the evaluation of a test, a new population is created for each such newly discovered targets. 
At the end of the search, MIO does output a test suite (i.e., a set of test cases) based on the best tests in the archive for each testing target.

For black-box testing, \evo employs a random test generation. 
The tool produces random inputs, but still syntactically valid with respect to the OpenAPI/Swagger schema of the SUT. 

Two different studies~\cite{kim2022arxiv,zhang2022arxiv} compared \evo with other fuzzers for RESTful APIs, showing that \evo gives the best results.
\evo is open-source, hosted on GitHub~\cite{EvoMaster}, with each release automatically published on Zenodo for long-term storage.
The extension presented in this paper is available to practitioners since \evo version $1.5.0$~\cite{andrea_arcuri_2022_6651631}.

%%%%%%%%%%%%%%%%%%%%%%%%%%%%%%%%%%%%%%%%%%%%%%%%%%%%%%%%%%%%%%%%%%%%%%%%%%%%
\section{Related Work}
\label{sec:relatedwork}

\subsection{Testing of GraphQL APIs}
Automated testing of GraphQL APIs is a topic that has been practically neglected in the research literature. 
To the best of our knowledge, so far only three approaches have been investigated regarding the automated testing of GraphQL APIs~\cite{vargas2018deviation,karlsson2020automatic,zetterlund2022harvesting} (besides the poster version of this extended paper~\cite{belhadi2022graphql}).

Vargas \etal~\cite{vargas2018deviation} proposed a technique called ``Deviation Testing''. 
It consists of three steps. 
In the first step, an already existing test case is taken as input. 
This test constitutes a base to seed and compare the newly generated tests. 
The second step is the test case variation, where variations of the initial seeded test case are generated using deviation rules (where a deviation consists of a small modification). 
Four types of deviation rules are defined: 
1) field deviation consists of adding and deleting the selection of fields in the original query; 
2) not null deviation consists of replacing a declared non null argument with null; 
3) type deviation consists of changing an argument type by another type;
4) empty fields deviation consists of deleting all fields and sub fields of the original query. 
The third step is the test case execution where the input test and its variations are executed. 
The last step consists of comparing the results between the input test and its variation (e.g., wrong inputs should lead to a response containing an error message). 

Karlsson \etal~\cite{karlsson2020automatic} proposed a black-box property based testing method. 
The method consists of the following steps. 
First, all specifications of the types and their relations are  extracted from the schema. 
Data is generated at random according to the schema, with customized ``data generators'' provided by the user.
In addition, the authors suggest two strategies to use as automated oracles: the first one aims to check the returned HTTP status codes, and the second one verifies that the resulting data returned conform to the given schema.

Zetterlund \etal~\cite{zetterlund2022harvesting} presented a technique to capture the HTTP calls done by users in production (e.g., when interacting with web frontend), and generate test cases for the GraphQL API in the backend. 
These generated tests can then be used for regression testing.

Our novel solution does not require any pre-existing test case (like in~\cite{vargas2018deviation}), nor it requires the user to write customized input generators (like in~\cite{karlsson2020automatic}), nor it requires users interacting with a frontend~\cite{zetterlund2022harvesting}.
Our framework benefits from the \evo tool, which is able to do advance white-box testing (e.g., using testability transformations~\cite{arcuri2020testability} and SQL interaction analysis~\cite{arcuri2020sql}), and also be able to perform  black-box testing when source code of the SUT is not available.
In this paper, we provide a complete testing pipeline, and an intelligent genetic-based exploration for the possible test case configurations.

\subsection{Testing of REST APIs}

In recent years, there has been an increasing interest in the research community about the automation of testing web services, where RESTful APIs are the currently the most common type.

For example, Godefroid et al.~\cite{godefroid2020differential} introduced the differential regression testing for avoiding breaking changes in the REST APIs. 
They analyzed both types of regressions in APIs, regression in the contract rules between the client and the server, and regression in the server itself, the different changes in the server versions.  
The differential testing is performed to automatically identify abnormal behavior in both kinds of regressions. 
It consists of comparing different versions of the server and also the different versions of the contracts with the client. 

Viglianisi et al.~\cite{viglianisi2020resttestgen} proposed an approach for automatically generating black-box test cases for REST APIs. 
It takes as input the Swagger/OpenAPI specification of the API and consists of three modules: 
It first analyzes the schema and computes its corresponding operation dependency graph which is a graph that represents the data dependencies. 
It then automatically generates test cases of the REST API by reading both the graph created in the previous step and the schema in order to test the nominal scenarios. 
It finally applies mutation operators to the nominal tests which violate data constraints for testing error scenarios. 
In order to decide whether a test is successful or not, two oracles are established based on the returned status codes and on the compliance with the schema. 

Lopez et al.~\cite{martin2021specification} presents a new formulation of the automated test APIs problem using  CSP (Constraint Satisfaction Problem).  
The  IDL (Inter-parameter Dependency Language) is first introduced to formally describe the different relations among input parameters of the REST API. 
Then, the CSP is used to automatically analyse the IDL specification.
Finally, a catalogue of analysis is constructed in order to extract helpful information such as checking whether an API call is valid or not

So far, all approaches presented in the literature for testing RESTful APIs are black-box.
\evo (Section~\ref{sec:evomaster}) is currently the only tool that can do both white-box and black-box testing. 
It uses evolutionary techniques for  white-box testing, and  random search for  black-box testing. 
Furthermore, recent comparisons of tools~\cite{Kim2022Rest,zhang2022open} show that \evo gives the best results on the selected APIs used in those tool comparisons.

%%%%%%%%%%%%%%%%%%%%%%%%%%%%%%%%%%%%%%%%%%%%%%%%%%%%%%%%%%%%%%%%%%%%%%%%%%%%%%%%%%%%%%%%%%%%%%%%%%%%%%%%%%%%%%%%%%%%
\section{GraphQL Test Generation}
\label{sec:EvoMaster+}

\begin{figure*}[htbp!]
	\centering
	\includegraphics[width=0.85\textwidth]{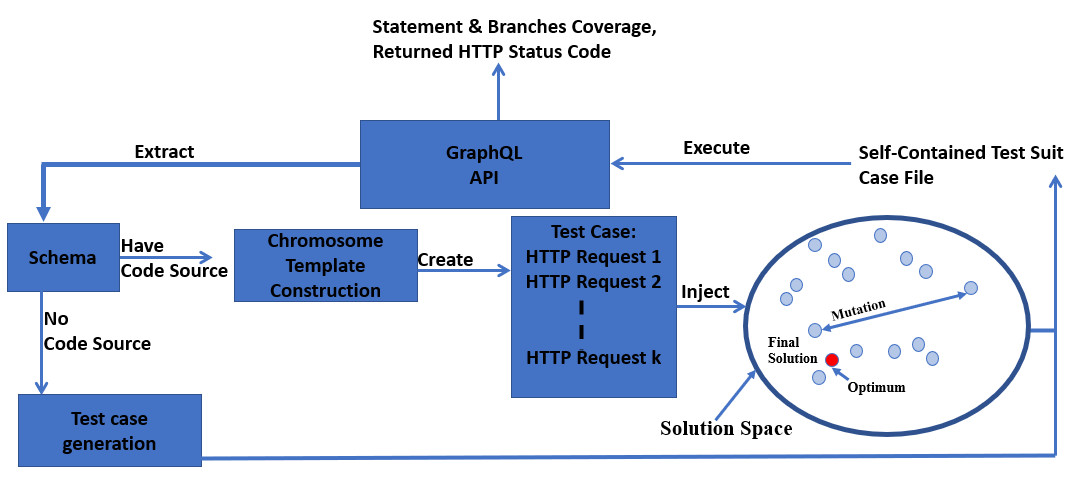}
	\caption{Framework for GraphQL test generation} \label{fig:EvoMaster+}
\end{figure*}

This section presents our novel proposed framework for automated test case generation of GraphQL APIs, built on top of the \evo tool. 
As sketched in Figure~\ref{fig:EvoMaster+}, the proposed framework targets both white-box and black-box testing. 
The white-box testing is performed when 
the information related to the schema is provided and have access to source code of the GraphQL API. 
In case of only having the schema, black-box testing is rather used.

\begin{algorithm}[ht!]
\caption{GraphQL Test Generation Algorithm}
 \label{algo:GraphQLTestGeneration}
\begin{algorithmic}[1]
\STATE \textbf{Input}: $G=\{N_1, N_2, \dots, N_n\}$: the set of $n$ data in the GraphQL API.
\STATE \textbf{Output}: Tests: the set of generated tests.
\STATE Schema $\leftarrow$ Extraction(G)
\IF{code source in G} 
\STATE Gene $\leftarrow$ ChromosomeTemplateConstruction(Schema)
\STATE TestCase $\leftarrow$ Create(Gene)
\STATE Tests $\leftarrow$ EvolutionaryAlgorithm(TestCase)
\ELSE 
\STATE Tests $\leftarrow$  Random(Schema)
\ENDIF
\RETURN Tests
\end{algorithmic}
\end{algorithm}

Algorithm~\ref{algo:GraphQLTestGeneration} presents the pseudo-code of the GraphQL test generation algorithm. 
The input data is the entry point to the GraphQL API (e.g., a URL).
In case of white-box testing, we also provide the  source code as input. 
The results will be a set of generated tests (e.g., in JUnit or Jest format). 
The process starts by extracting the schema from the graph data in line $3$. 
In case of having the  source code of the GraphQL API, the white-box solution is performed. 
It first determines the corresponding genes, creates the chromosome template, and applies an evolutionary algorithm process to generate the tests (from line $4$ to line $7$). 
In case of missing the source code of the GraphQL API, the black-box solution is established by generating random tests based on the schema, illustrated in line $9$. 
The algorithm returns the set of generated tests as shown in line $11$. 
In the following, both kinds of testing are explained in more details.

\subsection{White-Box Testing}
The process starts by extracting the schema from the GraphQL API. 
The chromosome template is then constructed from the schema. 
Test cases are represented by a sequence of HTTP requests, instantiated from the chromosome template. 
The test cases are evolved using MIO~\cite{mio2017}, where each test case will contain genes representing how to build the GraphQL queries based on the given schema. 
The evolutionary search is performed by applying two mutation operators to evolve the test cases (one to change the queries/mutations in each HTTP calls, and the other to add/remove HTTP calls in a test case). 
This enables to efficiently explore the solution space, with the aim of maximizing code coverage and fault finding. 
From the final evolved solution, a self-contained test suite file (e.g., in JUnit format) is generated as output of the search.

In the following, we describe the main components of the white-box testing in more details:
\begin{enumerate}
\item \textbf {Problem Representation}
In order to fetch the whole schema from a GraphQL API, an \emph{introspective} query is used. 
Given an entry point to the GraphQL API (e.g., typically a \texttt{/graphql} HTTP endpoint), GraphQL enables a standard way to fetch a schema description of the API itself. 
The schema specifies all the information about the available operation types, such as queries, mutations and all available data types on each of them. 
As a result, the GraphQL schema is returned in JSON format. 
This latter is then parsed in our \evo extension and used to create a set of action templates, one for each query and mutation operation. 
Each action will contain information on the fields related to input arguments (if any is present) and return values. 
A chromosome template is defined for each action, which is composed of non-mutable information (e.g., field's names) and a set of mutable genes. 
In this context, each gene characterizes either an argument or a return value in the GraphQL query/mutation.
For objects as return values, a query/mutation must specify which fields should be returned (at least one must be selected), and so on recursively if any of the selected fields are objects as well.  
To represent the fact that a field is always optional for queries, a return gene is modeled by an object gene where all its fields are optional. 
However, we had to extend the mutation operator in \evo with a post-processing phase, to guarantee that at least one field gene is selected during the search.
In other words, if after a mutation of a gene, which represents a returned object value in the GraphQL query/mutation,  all fields are de-selected, then the post-processing will force the selection of one of them (and so on recursively if the selected field is an object itself).  
On the other hand, if a return value is a primitive type, then there is no need to create any gene for it, as there is no selection to make. 
Furthermore, similar to functions calls, fields in the returned value can have input argument themselves.
When a returned value for a parent field is executed, both input arguments and returned value are recursively selected to generate a child field value until it produces a scalar value whether in input arguments or in returned values.
To model those function calls we introduced a new special type of gene called \texttt{Tuple}, discussed next.

To fully represent what is available from the GraphQL specification, the following kinds of gene types from \evo have been re-used and adapted:
\begin{enumerate}
\item \texttt{String}: It contains string variables which are defined by an array of characters. 
A minimum length of the string is zero which represents the empty string. 
Each string gene cannot exceed a predefined maximum number of characters (e.g., 100).

\item \texttt{Enum}: This gene represents the enumeration type, where a set of possible values is defined, and only one value is activated at a given time. 
The elements in the set can be in different formats (e.g., enumerations of numbers or enumerations of strings).

\item \texttt{Float/Integer/Boolean}: genes representing variables with simple data types. 
Boolean genes represent variables with true or false values. 
Integer and float genes represent integer and real-value variables, respectively.

\item \texttt{Array}: This gene represents a sequence of genes with the same type. 
This gene has variable length, where elements can be added and removed throughout the search.
In order to mitigate creating too large test cases, for instance with millions of genes, the size of an array gene should not exceed a given threshold.

\item \texttt{Object}: 
This gene defines an object with a specific set of internal fields. 
Differently from the array gene, where the elements should be with the same type, an object gene may contain elements with different types. 
To do so, this gene is represented by a map, where each key in the map is determined by the field name in each element in the object.

\item \texttt{Optional}: A gene containing another gene, whose presence in the phenotype is controlled by
a boolean value. 
This is needed for example to represent nullable types in arguments and selection of fields in returned objects.

\item \texttt{CycleObject}: This special gene is used as a placeholder to avoid infinite cycles, when selecting object fields that are objects themselves, which could be references back to the starting queried object. 
Once a test case is sampled, its gene tree-structure is scanned, and all \texttt{CycleObject} genes are forced to be excluded from the phenotype (e.g., if inside an \texttt{Optional} gene, that gets marked as non-selected, and the mutation is prevented to select it; if the \texttt{CycleObject} is the type for an Array gene, such array gets a fixed size of 0, and the mutation operator is prevented from adding new elements in it). 

\item \texttt{LimitObject}: GraphQL schemas are often very large and complex, and the levels of nesting fields can be potentially huge. 
We use this special gene as a placeholder when a customized depth limit is reached. 
The depth is the number of nesting levels of the object fields.

\item \texttt{Tuple}:  This gene is needed for example when representing the inputs of function calls.
It is composed of a list of elements of possible different types, where the last element can be treated specially. 
This is for example the case of function calls when the return type is an object, on which we need to select what to retrieve (and these selected elements could be function calls as well, and so on this is handled recursively).

\end{enumerate}

After defining the possible type of genes supported by the proposed framework, we consider the solution space, where each solution is a set of test cases. 
A test case is composed of one or more HTTP request.  
In order to represent an HTTP request, we typically need to deal with its components: HTTP verb, path and query parameters,  body payloads (if any) and headers.  

A GraphQL request can be sent via HTTP GET (used only for queries) or HTTP POST methods with a JSON body (used for queries and mutations).
For simplicity, we only use the verb POST for both queries and mutations. 
A GraphQL server uses a single URL endpoint (typically \texttt{/graphql}), where the HTTP requests with the GraphQL queries/mutations will be sent.
In the context of test generation for a GraphQL API, the main decisions to make are on how to create JSON body payloads to send.   
The genotype will contain genes (from the set defined above) to represent and evolve such JSON objects.

\item \textbf {Search Operators and Fitness Function}

Once a chromosome representation is defined based on the GraphQL schema, test cases are evolved and evaluated in the same way as done for RESTful APIs in \evo (recall Section~\ref{sec:evomaster}), including testability transformations~\cite{arcuri2020testability} and SQL database handling~\cite{arcuri2020sql}.
Internally, the MIO algorithm is implemented in a generic way, independently of the addressed problem (e.g., REST and GraphQL APIs), and it is only a matter of defining an appropriate phenotype mapping function (e.g., how to create a valid HTTP request for a GraphQL API based on the evolved chromosome genotype). 

When evaluating the fitness of an evolved test, besides considering testing targets related to code coverage and HTTP status coverage (for each different query/mutation operation), we also create new testing targets based on the returned responses.
As discussed in Section~\ref{sub:GraphQl}, each response could contain either a \texttt{data} field or an \texttt{errors} field.
For each query and mutation in the GraphQL schema, we consider two additional testing targets for those two possible outcomes.
Note that a trivial way to get a response with \texttt{errors} is to send a syntactically invalid query.
As such evolved test cases would be of little use, we explicitly avoid generating such kind of test cases.

As a given query/mutation might fail for different reasons, we keep track of the last executed line in the business logic of the SUT.
We further create a separated testing target for each combination of errored query/mutation and last executed line. 
Having explicit testing targets for those cases enables MIO to save in its archive such evolved test cases, albeit the fitness function would have (currently) no gradient to lead to generate such kind of test cases in the first place.  

\end{enumerate}

\subsection{Black-Box Testing}
We use the black-box testing when we do not have any knowledge about the source code of the GraphQL API, or it is not available for instrumentation (e.g., to calculate the search-based heuristics like the branch distance). 
It is not straightforward to get a high coverage value for such kind of tests, as little information from the SUT can be exploited.
However, in some cases (e.g., when testing remote services), a black-box approach might be the only option available for automated testing. 
We use the random search algorithm developed in \evo. 
The main idea behind random search is performing a randomized process in generating the test cases,
where no fitness function is employed. 
The reason of not using search-based heuristics is due to the lack of the  source code of the GraphQL APIs. 

Using the same process to sample new test cases for white-box testing, in the random search we sample a series
of test cases, which is saved in an archive. 
As no code information is available, test cases are added in the archive only if they cover new targets based on the HTTP responses, based on each query/mutation operation in the schema (in the same way as we do for white-box testing).
In other words, for each query/mutation we retain test cases that lead to different HTTP status codes, and at least one with a correct \texttt{data} response and at least one with an \texttt{errors} response.

The main steps of the random search can be summarized as follows: 

\begin{enumerate}
    \item Test case generation: The test cases are generated in a random way, but they are still syntactically valid. 
    For instance, if we consider the example illustrated in Figure~\ref{fig:DataGraph}, the test cases are generated by exploring the fields of the pets node. 
    For instance, if we consider the field ``id'' an integer represented in 32 bits. 
    The possible test cases for the field ``id'' is $2^{32}$. 
    We also explore different combinations of two or more fields in each node. 
    For instance, consider the same example illustrated in Figure~\ref{fig:DataGraph}, the test cases might be generated from both fields ``id'', and ``name'' of the node pets. 
    If we consider the length of the string is limited to $10$, the possible tests cases for the field ``name'' is $2^{160}$ (assuming each character being 2-bytes). 
    Therefore, the number of possible test cases by only exploring the fields ``id'', and ``name'' is $2^{32} \times 2^{160}$, which results an immense search space. 
    Therefore, in our implementation, and in order to mitigate the combinatorial explosion, we use threshold to limit the number of generated test cases (i.e., we limit the number of test cases we sample during the random search).  
    \item Fault determination: After a test case is sampled, the test is evaluated by calling the GraphQL endpoint, to detect possible faults (e.g., based on HTTP status codes and \texttt{errors} fields in the responses). 
\end{enumerate}

%%%%%%%%%%%%%%%%%%%%%%%%%%%%%%%%%%%%%%%%%%%%%%%%%%%%%%%%%%%%%%%%%%%%%%%%%%%%%%%%%%%%%%%%%%%%%%%%%%%%%%%%%%%%%%%%%%%%
\section{Empirical Study}
\label{sec:Empirical Study}
\subsection{Experimental Setup}

In this section, several experiments have been carried out to validate the applicability of the proposed framework for GraphQL test generation. 
This can be achieved by answering the three following research questions:

\begin{description}

\item[{\bf RQ1}:] For white-box testing of GraphQL APIs, how effective is MIO at maximizing code coverage and fault detection compared to Random search?

\item[{\bf RQ2}:] How does black-box testing fare on existing APIs on the Internet?

\item[{\bf RQ3}:] What kinds of faults are found by our novel technique? 
	
\end{description}

%---------------------------------------------------------------------------------------
\subsubsection{White-Box}
GitHub~\cite{Github}, arguably the main repository for open-source projects, was used to find SUTs for experimentation. 
JVM and NodeJS projects were scanned and filtered, while excluding trivial projects. 
For this study, seven GraphQL web services are selected, which we could compile and run with no problems:

\begin{itemize}
    
\item The Spring \textit{petclinic}~\cite{gqlpetclinic} API (4 567 LOCs) is an animal clinic where a pet owner can register his pet for an examination.
The examination is carried out by a veterinarian who has one or more specialist areas.

\item \textit{patio-api}~\cite{gqlpatio} (12 552 LOCs) is a web application that attempts to estimate the happiness of a given team periodically by asking for a level of happiness.

\item \textit{graphql-ncs}(548 LOCs) and \textit{graphql-scs} (577 LOCs) are based on  artificial RESTfull APIs from an existing benchmark~\citep{EMB}. For this study, we adapted these two APIs into GraphQL APIs. 
\textit{graphql-ncs} and \textit{graphql-scs} are based on a code that was designed for studying unit testing approaches on solving numerical~\citep{ArB11a} and string~\citep{Alshraideh06} problems.

\item \textit{react-finland}~\cite{gqlreact-finland} (16 206 LOCs) is an API for a week long developer conference focused on React.js and related technologies. 

\item \textit{timbuctoo}~\cite{gqltimbuctoo} (85 365 LOCs) is an API that allows scientists to decide how data from different databases is shared.

\item \textit{e-commerce}~\cite{gqlecommerce} (1 791 LOCs) an e-commerce API built on Phoenix and Elixir that can be utilized  in order to  create  interactive e-commerce web applications.

\end{itemize}

To the best of our knowledge, there is no other existing white-box fuzzer that can be used to test GraphQL APIs.
Therefore, in this paper we cannot compare with any existing technique, as none is available.
White-box fuzzing GraphQL APIs is a novel contribution of this paper.
Still, it is important to verify whether a novel sophisticated technique is really warranted, and no simpler technique would be already as effective~\cite{ali2009systematic}.
When nothing else is available, 
a common baseline in software testing research is \emph{Random Testing}~\cite{AIB11}, in which an application is tested with random inputs.
Still, sending random bytes on the TCP connection the SUT is listening on would be of little to no value, as the chances of generating a valid GraphQL query (or even simply a valid HTTP request) would be virtually non-existent. 
Therefore, for doing random testing, we still sample and send syntactically valid GraphQL queries based the schema of the SUT.

Our proposed framework is integrated in \evo, where a comparison between MIO and the baseline random search algorithm is carried out.
For the experiments, we set 100 000 HTTP calls as search budget for our white-box testing approach.
To take into account the randomness of the algorithm, each experiment was repeated 30 times~\cite{Hitchhiker14}. 
We selected covered testing targets (\#Targets), line coverage (\%Lines), and the number of detected faults (\#Errors) as metrics for comparisons. 
The testing target (\#Targets) is the default coverage criterion in \evo. 
It comprises and aggregates different metrics, such as code coverage (including branch coverage), HTTP status code coverage and fault findings.
The line coverage (\%Lines) is collected as part of our code instrumentation.
Furthermore, we also reported (\#Errors) by identifying potential faults, i.e., 500 HTTP status codes and responses with \texttt{errors} entries (recall Section~\ref{sec:EvoMaster+}).

%---------------------------------------------------------------------------------------------
\subsubsection{Black-Box}

\begin{table*}[t!]
\centering
\scriptsize
\caption{GraphQL APIs used for black-box experiments} \label{tab:gqlbapis}
\begin{tabular}{llc}
\hline
 Name & Description &  Source Code\\ 
\hline
\emph{aniList}& provides access to anime and manage entries, including character, staff, and & Yes\\
&live airing data. & \\
\emph{deutsche-bahn}& infrastructure Data, like realtime facility status and stations & Yes\\
\emph{barcelona-urban-mobility}& combine information about the different urban mobility services of Barcelona city  & Yes \\
\emph{buildkite}& a platform for continuous integration and deployments & No \\
\emph{câmara-dos-deputados}& an api to obtain the data from the brazilian deputies chamber & Yes \\
\emph{catalysis-hub} & chemical surface reaction energies and structures & Yes\\
\emph{contentful}& provides a content infrastructure for digital teams to power content in websites, & Yes \\ &apps, and devices & \\
\emph{countries} & information about countries, continents and languages &Yes \\
\emph{demotivational-quotes}& get random demotivational quote & Yes \\
\emph{digitransit}&  journey planning solution combining several open source components into available & No\\ 
&route planning service & \\
\emph{ehri}& holocaust-related archival materials held in institutions across Europe and beyond & No \\
\emph{fauna}& serverless GraphQL database  & No \\
\emph{fake-graphQL-api}& mock user and to do data & No \\
\emph{fruits}& provides information of fruit trees of the world  & Yes \\
\emph{ghibli}& catalogs the people, places, and things found in the worlds of Ghibli & Yes \\
\emph{gitLab}& host-your-own Git repository hosting service  & No \\
\emph{google-directions}& GraphQL wrapper over the google directions API & Yes \\
\emph{music-brainz}& an open music encyclopedia that collects music metadata & Yes \\
\emph{pokémon}& query for all the Pokémon data including their abilities, moves, items, learnsets,  & Yes \\
&and type matchups & \\
\emph{hivdb}& a curated database to represent, store and analyze HIV drug resistance data & No \\
\emph{jobs}& GraphQL jobs directory & No\\
\emph{melody}& fast and reliable dependency manager for Go programming language & No\\
\emph{react-finland} & GraphQL API for conferences and meetings & Yes \\
\emph{rickandmortyapi}& based on the television show Rick and Morty. Provide the Rick and Morty & Yes \\
&information (characters, episodes, locations) & \\
\emph{spacex} & a non official platform for SpaceX’s data & No \\
\emph{spotify}& provides instant access to millions of songs, from old favorites to the latest hits & Yes \\
\emph{swapi}& provides all the Star Wars data & Yes  \\
\emph{swop}& GraphQL foreign exchange rate API & No  \\
\emph{travelgateX}& a global marketplace for the travel trade & No \\
\emph{universe}& check what your friends are doing and find unique events near you using our filter & No \\
\emph{weather}& retrieve the current weather for any given city & Yes \\
\hline
\end{tabular}
\end{table*}

To evaluate the black-box testing, 31 online APIs with different domain applications, and different number of endpoints have been selected from \emph{apis.guru}~\cite{gqlapisguru}, a curated public listing of available web services on Internet. 
These APIs are written with different programming languages, such as JavaScript and Python. 
Some APIs provide their implementation (e.g., open-source), whereas others do not (e.g., commercial services). 
When an API required authentication, we created an account on these APIs, and added the right authentication information the HTTP headers of \evo (e.g., authentication header can be set with the command-line argument \texttt{--header}). 
For our experiments, we considered all the GraphQL APIs listed on \emph{apis.guru},
but excluded APIs that are no longer available (but still listed on \emph{apis.guru}), or that required payment to create an account. 
Table~\ref{tab:gqlbapis} gives short description of the 31 APIs used in our experiments. 

We ran our extension of \evo on all those APIs with black-box mode. 
The stopping criterion was set to 1 000 HTTP calls per run.
Each experiment was run only three times, since sending thousands and thousands of HTTP calls to live services could be interpreted as Denial-of-Service attack.
For the same reason, we put a rate limiter of at most 10 HTTP requests per minute (i.e., \evo would make HTTP calls only every 6 seconds).

As we do not have any control on these remote APIs, repeating the experiments more times would not add much more information, as such runs would not be fully independent.

%---------------------------------------------------------------------------------------
\subsection{Experiment Results}

\begin{table}
	%\small
	\centering
	\caption{ Results for 100k HTTP call budget for white-box testing}
	\label{tab:rq1-100k}
	\resizebox{0.9\textwidth}{!}{
		\begin{tabular}{  l rrrr  rrrr rrrr }\\ 
\toprule 
SUT & \multicolumn{4}{c}{Targets \#} & \multicolumn{4}{c}{Line Coverage \%} & \multicolumn{4}{c}{Errors \#} \\ 
    & RS & WB  & $\hat{A}_{12}$ & p-value  & RS & WB  & $\hat{A}_{12}$ & p-value  & RS & WB  & $\hat{A}_{12}$ & p-value \\ 
\midrule 
\emph{ecommerce-server} & 340.0 & 340.1 & 0.58 & 0.218 & 8.2 & 8.0 & {\bf 0.08} & $< 0.001$ & 28.0 & 27.6 & {\bf 0.37} & 0.029 \\ 
\midrule 
\emph{graphql-ncs} & 365.8 & 552.1 & {\bf 1.00} & $< 0.001$ & 51.4 & 78.3 & {\bf 1.00} & $< 0.001$ & 6.0 & 8.2 & {\bf 1.00} & $< 0.001$ \\ 
\midrule 
\emph{graphql-scs} & 634.3 & 723.3 & {\bf 1.00} & $< 0.001$ & 68.1 & 75.7 & {\bf 1.00} & $< 0.001$ & 0.0 & 0.8 & {\bf 0.79} & $< 0.001$ \\ 
\midrule 
\emph{patio-api} & 3538.4 & 3754.6 & {\bf 1.00} & $< 0.001$ & 39.0 & 43.9 & {\bf 0.97} & $< 0.001$ & 59.5 & 85.2 & {\bf 1.00} & $< 0.001$ \\ 
\midrule 
\emph{petclinic} & 652.4 & 650.3 & 0.43 & 0.328 & 60.0 & 59.7 & {\bf 0.40} & 0.011 & 17.7 & 17.3 & {\bf 0.28} & 0.001 \\ 
\midrule 
\emph{react-finland} & 315.2 & 529.0 & {\bf 1.00} & $< 0.001$ & 1.0 & 1.8 & {\bf 1.00} & $< 0.001$ & 24.5 & 30.5 & {\bf 1.00} & $< 0.001$ \\ 
\midrule 
\emph{timbuctoo} & 6787.2 & 7395.9 & {\bf 1.00} & $< 0.001$ & 27.3 & 29.3 & {\bf 0.99} & $< 0.001$ & 43.9 & 87.4 & {\bf 0.99} & $< 0.001$ \\ 
\midrule 
Average  & 1804.8 & 1992.2 & 0.86 &  & 36.4 & 42.4 & 0.78 &  & 25.7 & 36.7 & 0.77 &  \\ 
\bottomrule 
\end{tabular} 

	}
\end{table}

%-----------------------------------------------------------------------------------------
\subsubsection{Results for RQ1}

To compare MIO with Random, Table~\ref{tab:rq1-100k} reports their average \#Targets, \%Lines, \#Errors and an analysis of the pairwise comparisons using Mann-Whitney-Wilcoxon U-tests ($p$-$value$) and Vargha-Delaney effect sizes ($\hat{A}_{12}$), for each of the case studies.

When looking at the achieved target coverage, there is a clear improvement of white-box evolutionary search (i.e., MIO) compared to random search.
On 5 out of 7 APIs, the effect-size is maximum, i.e., $\hat{A_{12}}=1$.
This means that, in \emph{each} of the 30 runs of MIO, the results were better than the best run of random search.
In absolute terms, the largest \#Target improvement for MIO is on \textit{timbuctoo} (i.e., +608.7 covered targets than Random), which is the largest of the SUT used in our study.
Relatively, it is a +8.9\% improvement (from 6787.2 to 7395.9).
However, the largest relative improvement is for \emph{react-finland}, 
which is +67.8\% (from 315.2 to 529.0).
For \emph{ecommerce-server} and \emph{petclinic} there is no statistically significant difference (i.e., $p$-values are greater than the $\alpha=0.05$ threshold).

When looking at only line coverage, 
we can see that MIO enables covering 78.3\% of lines in \textit{graphql-ncs}.
This is a large improvement compared to the 51.4\% of Random (i.e., +26.9\%).
On average among the SUTs, the improvement is +6\% (i.e., from 36.4\% to 42.4\%).
When looking at the absolute  values for line coverage, we need to point out an issue with collecting coverage for NodeJS applications (i.e., \emph{ecommerce-server} and \emph{react-finland}), as coverage computation is not considering what achieved during boot-time of the API.

When looking at line coverage along, there is statistically worse results for \textit{petclinic} and  \textit{e-commerce}.
However, the differences are minimal (i.e., at most -0.3\%).
 For instance,  \textit{petclinic} is a  simple API used for demonstration, where large parts of its code is not executed (e.g., it has 3 different implementations of its data-layer, where only one can be active at a time, and this has to be specified in a configuration file when the API is started).  
On simple problems, random search can be already very good, whereas evolutionary search can have some small side-effects (which would likely disappear when using a longer search budget).
This is particularly the case if the fitness function does not provide gradient to the search.

In Table~\ref{tab:rq1-100k}, we also report the number of potential faults identified by both MIO and Random.
The table shows clear better results for MIO compared to Random, with an average high effect size of $\hat{A}_{12} =0.77$.
On \textit{timbuctoo}, MIO achieves the most, finding 87.4 errors on average compared to 43.9 for Random. 
This result is achieved thanks to the evolutionary operators adopted in the proposed framework to handle GraphQL APIs.

On our employed hardware, 100k calls take for example roughly 1 hour for \textit{petclinic}, and 2 hours for \textit{patio-api}.
In this context, an automatically achieved coverage of 43.9\% for a complex API like \textit{pation-api} could be considered a good, practical result, although more still need to be done (e.g., better search heuristics).
Not only many potential faults are found, but the generated tests can also be used for regression testing (i.e., they can be added to the test suites of the SUTs, and run as part of Continuous Integration to check if any change is breaking any current functionality).

\begin{figure}[t!]
	\centering
	\begin{subfigure}{.3\textwidth}
			\includegraphics[width=1\textwidth]{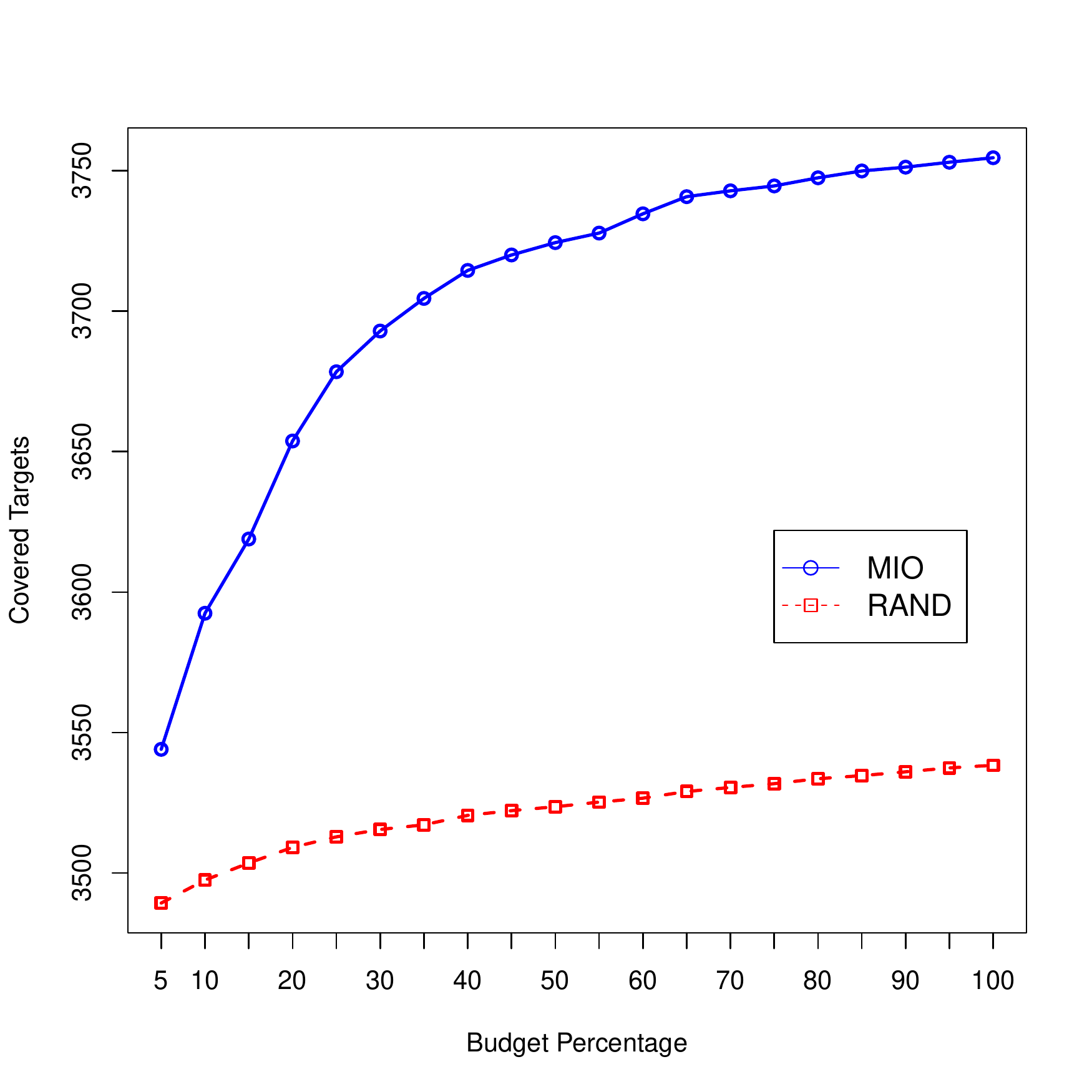}
			\caption{patio-api}
			\label{fig:patio}
	\end{subfigure}
	\begin{subfigure}{.3\textwidth}
		\includegraphics[width=1\textwidth]{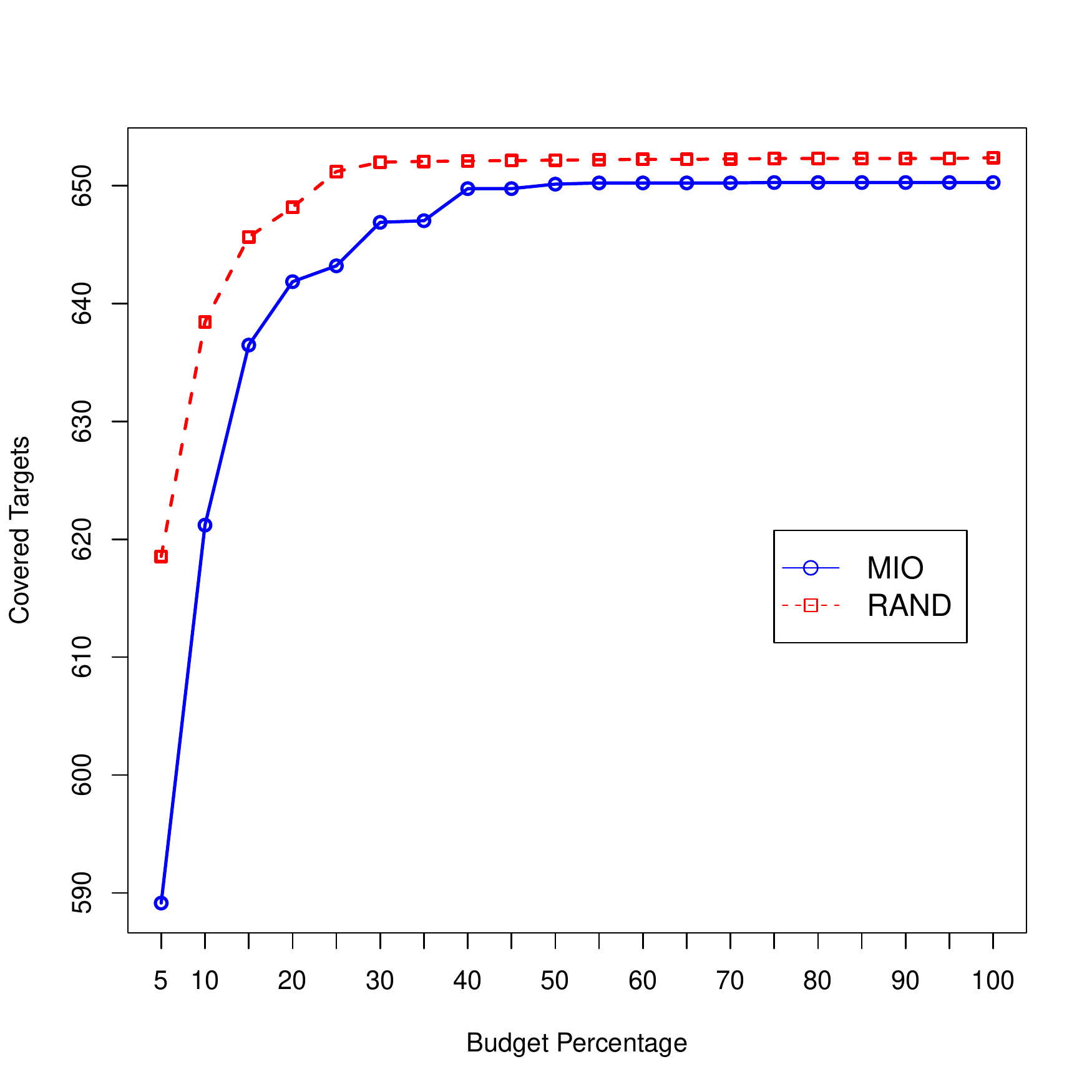}
		\caption{petclinic}
		\label{fig:petclinic}
	\end{subfigure}
	\begin{subfigure}{.3\textwidth}
			\includegraphics[width=1\textwidth]{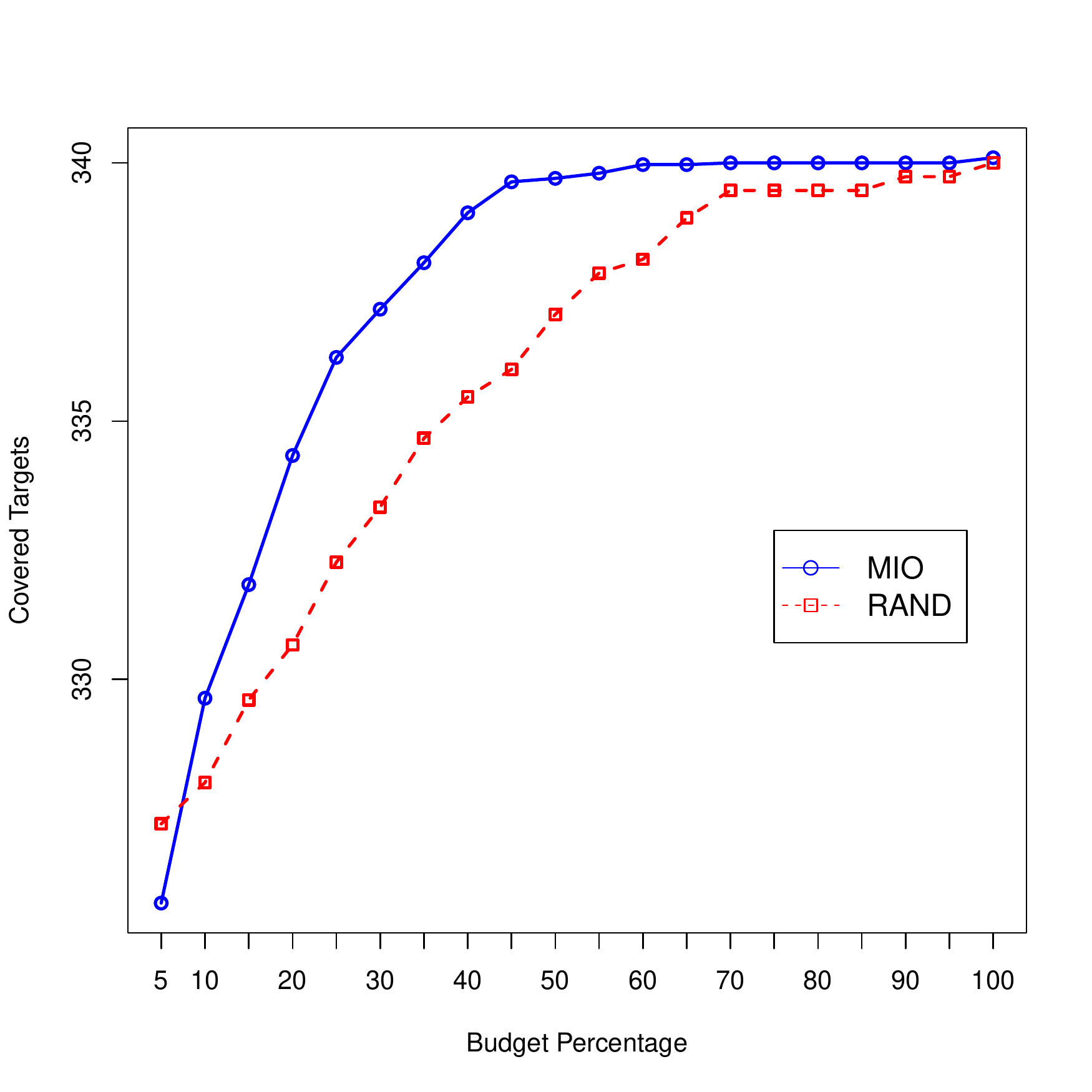}
			\caption{ecommerce-server}
			\label{fig:patio}
	\end{subfigure}\hfill
	\begin{subfigure}{.3\textwidth}
		\includegraphics[width=1\textwidth]{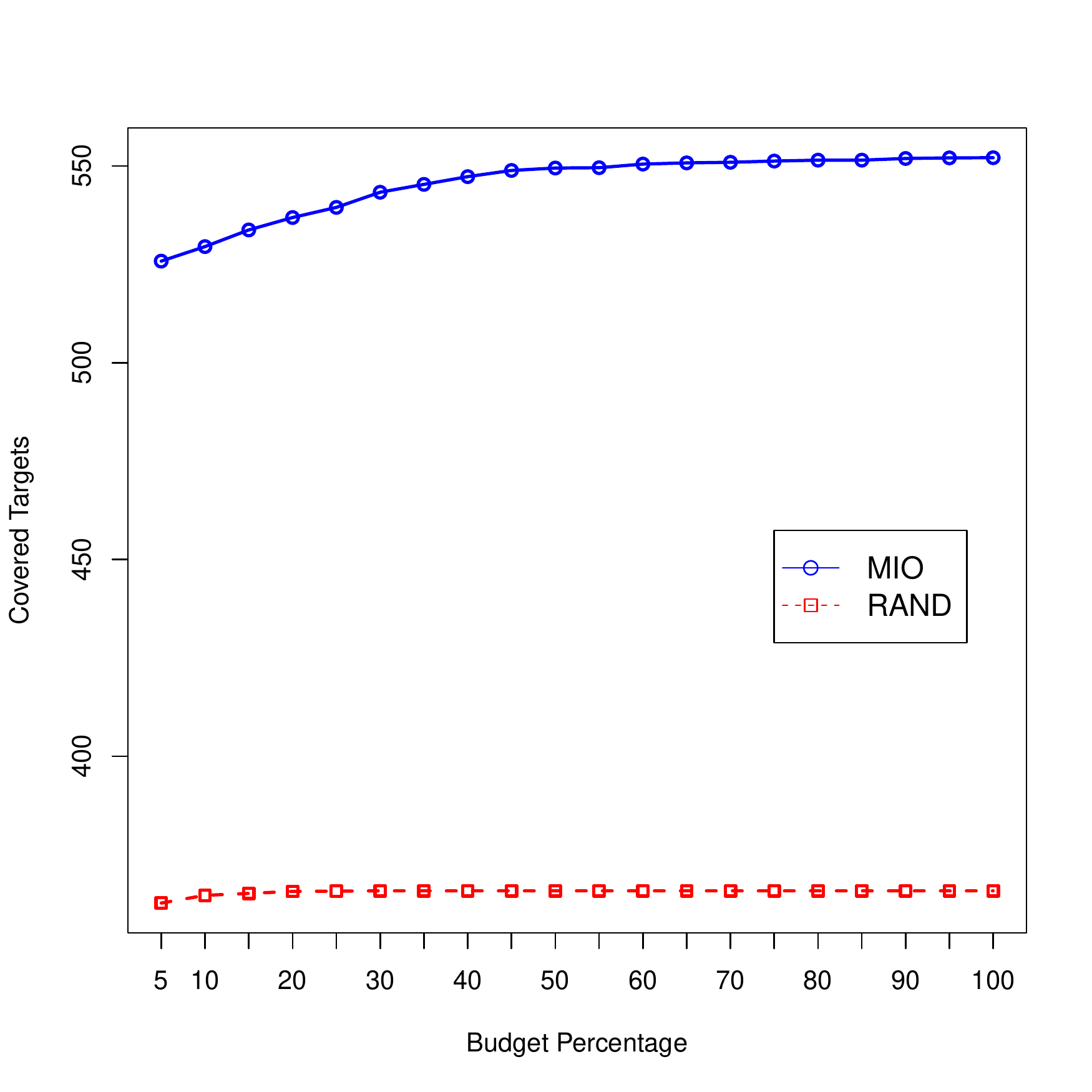}
		\caption{graphql-ncs}
		\label{fig:petclinic}
	\end{subfigure}
	\begin{subfigure}{.3\textwidth}
			\includegraphics[width=1\textwidth]{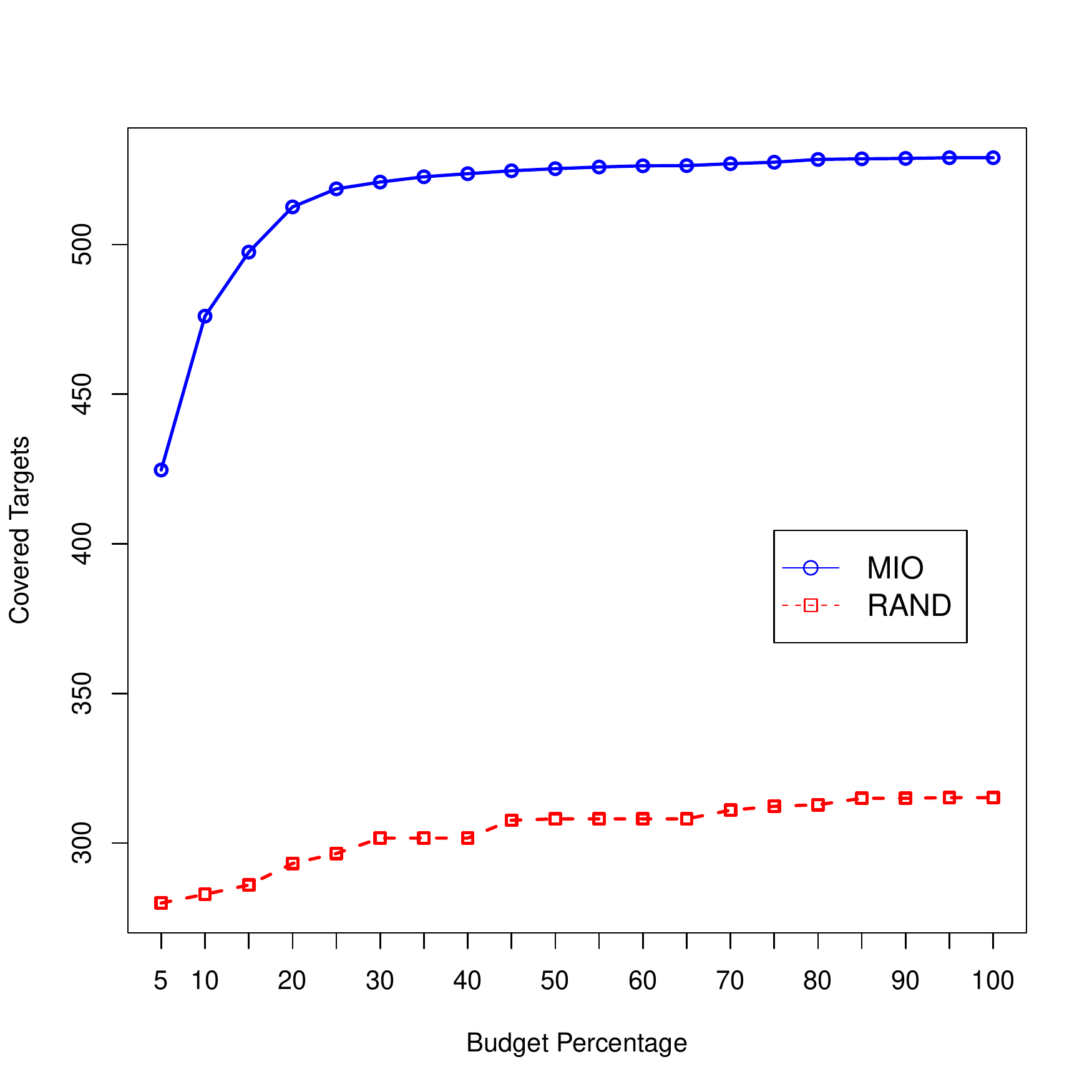}
			\caption{react-finland}
			\label{fig:patio}
	\end{subfigure}
	\begin{subfigure}{.3\textwidth}
		\includegraphics[width=1\textwidth]{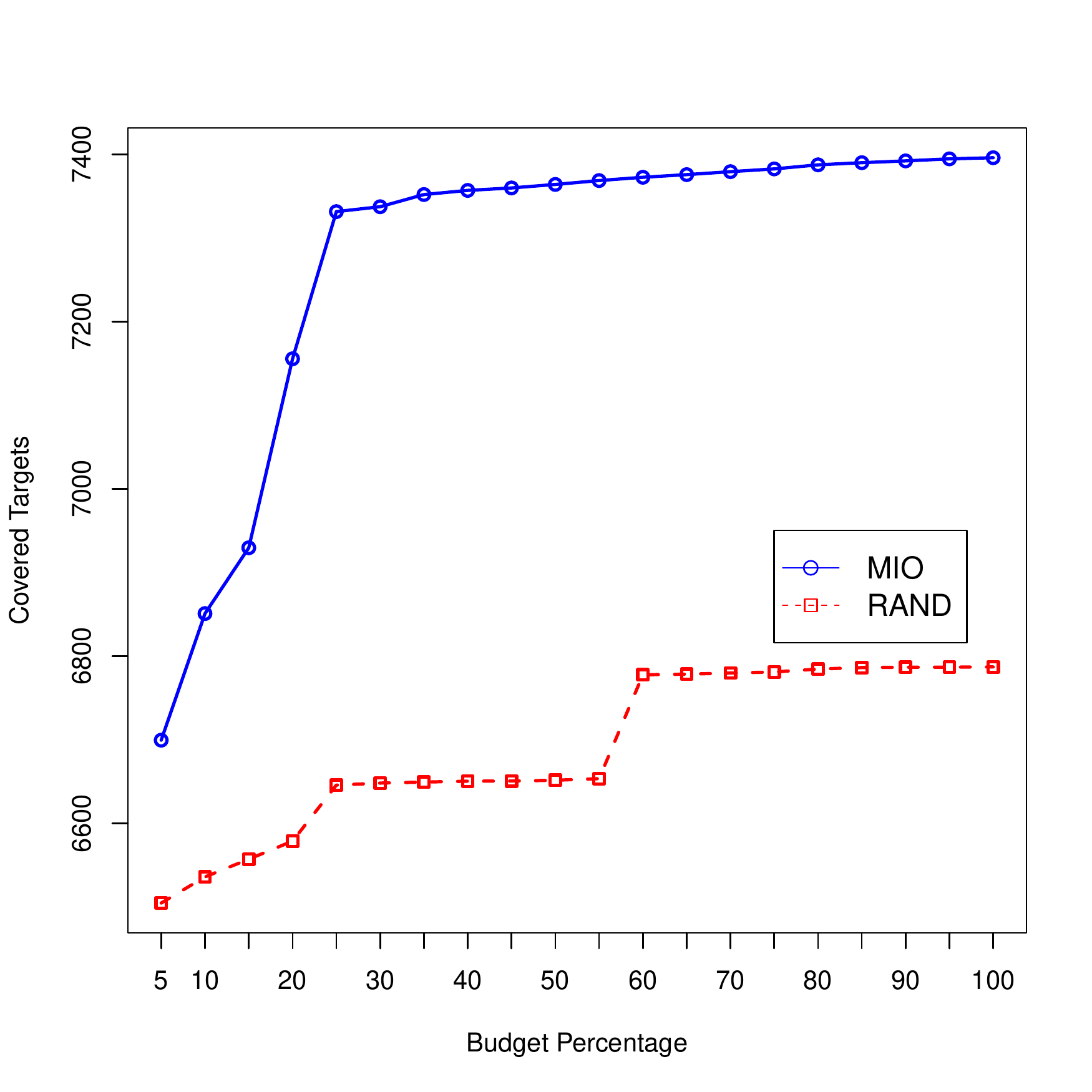}
		\caption{timbuctoo}
		\label{fig:petclinic}
	\end{subfigure}
	\begin{subfigure}{.3\textwidth}
			\includegraphics[width=1\textwidth]{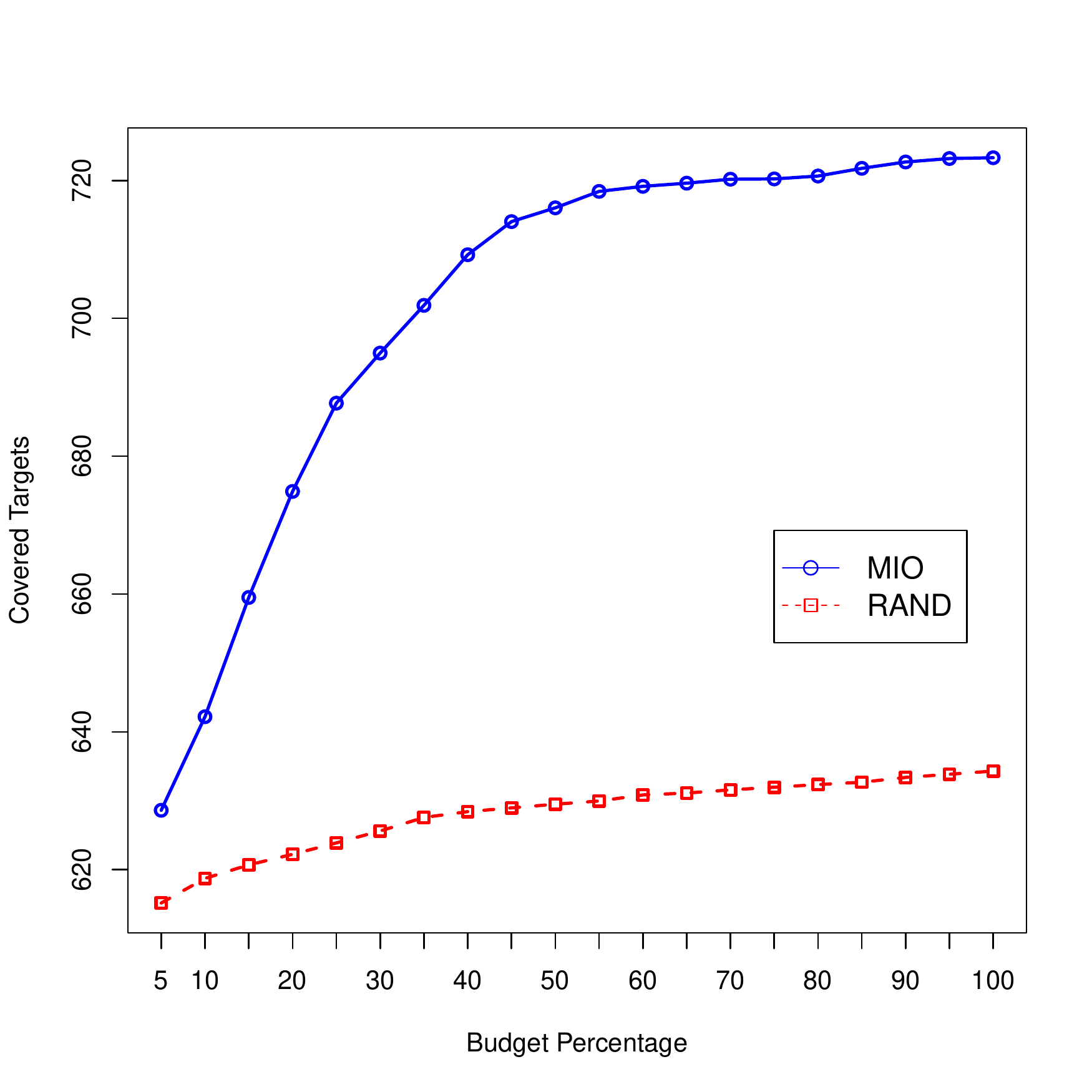}
			\caption{graphql-scs}
			\label{fig:patio}
	\end{subfigure}
	\caption{Covered targets throughout the search }
	\label{fig:plot}
\end{figure}

The choice of using 100k HTTP calls as stopping criterion is technically arbitrary.
It could had been more, or less. 
Such choice was based on what practitioners could use in practice~\cite{meituanArxiv2022}, although they would likely use ``time'' as stopping criterion (e.g., run the fuzzing for 10 minutes or 1 hour).
Using the number of maximum HTTP calls is done for scientific reasons, as it makes the experiments easier to replicate (e.g., they are not particularly impacted by the used hardware).
Nevertheless, the chosen search budget can impact the conclusions taken from the comparisons of search algorithms.
For this reason, in Figure~\ref{fig:plot} we report plot-lines for demonstrating the performances of the two compared techniques for the number of covered targets throughout the search, collected at each 5\% intervals (i.e., at each 5 000 HTTP calls). 

According to the reported results, MIO outperforms Random for all cases except for \textit{petclinic}, where we observed a slight gain for Random. 
The improvements of MIO are visible throughout the entire  search.
In a few cases, already with small budgets (e.g., 5 000 calls) MIO gets better results than Random at 100 000 calls. 
One interesting case to discuss is \emph{ecommerce-server}. 
For small budget, MIO is worse.
Then, for most of the search, it gives significantly better results than Random.
However, by the end of the search, the performance of the two algorithms converges. 
Given infinite time, even a trivial random search can achieve full coverage of the feasible targets~\cite{AIB11}.
Therefore, based on the chosen search budget, one could reach very different conclusions.
Using a budget somehow representing how practitioners would use these fuzzers in practice is therefore essential.

\begin{result}
	\textbf{RQ1}: In terms of covered targets, MIO demonstrates a consistent and significant improvements (+6\% line coverage and +11 more faults found on average) compared with  random testing. 
	This shows the effectiveness of MIO adapted for GraphQL testing for maximizing code coverage and fault detection.
\end{result}

%-------------------------------------------------------------------------------------------------
\subsubsection{Results for RQ2}

\begin{table}
	%\small
	\scriptsize
	\centering
	\caption{ Results for black-box testing}
	\label{tab:bb}
	%\resizebox{0.9\textwidth}{!}{
		\begin{tabular}{ l r r r}\\ 
\toprule 
SUT & \#Endpoints & \%NoErrors & \%WithErrors \\ 
\midrule 
\emph{Anilist}&  56&  1.8&  98.2\\ 
\emph{Bahnql}&  7&  100.0&  100.0\\ 
\emph{barcelona-urban-mobility}&  10&  50.0&  100.0\\ 
\emph{Buildkite}&  70&  7.6&  99.0\\ 
\emph{Camara-deputados}&  33&  23.2&  100.0\\ 
\emph{Catalysis-hub}&  11&  100.0&  100.0\\ 
\emph{Contentful}&  23&  11.6&  100.0\\ 
\emph{Countries}&  8&  87.5&  12.5\\ 
\emph{Demotivation-quotes}&  2&  100.0&  0.0\\ 
\emph{Digitransit}&  33&  26.3&  93.9\\ 
\emph{Directions}&  6&  33.3&  83.3\\ 
\emph{Ehri}&  19&  73.7&  100.0\\ 
\emph{Fauna}&  13&  10.3&  100.0\\ 
\emph{Fruits}&  7&  85.7&  28.6\\ 
\emph{Ghibliql}&  10&  100.0&  30.0\\ 
\emph{Gitlab}&  270&  1.9&  95.2\\ 
\emph{Graphbrainz}&  6&  0.0&  100.0\\ 
\emph{Graphqlpokemon}&  13&  53.8&  76.9\\ 
\emph{Hivdb}&  9&  44.4&  77.8\\ 
\emph{Jobs}&  15&  13.3&  86.7\\ 
\emph{Melody}&  2&  16.7&  100.0\\ 
\emph{Mocki}&  4&  100.0&  0.0\\ 
\emph{React-finland}&  13&  35.9&  100.0\\ 
\emph{RickAndMortyapi}&  9&  63.0&  66.7\\ 
\emph{Spacex}&  43&  93.8&  68.2\\ 
\emph{Spotify}&  11&  0.0&  100.0\\ 
\emph{Swapi}&  13&  46.2&  100.0\\ 
\emph{Swop}&  6&  16.7&  100.0\\ 
\emph{Travelgatex}&  11&  100.0&  0.0\\ 
\emph{Universe}&  90&  10.4&  94.4\\ 
\emph{Weather}&  2&  100.0&  100.0\\ 
\midrule 
\emph{Total} &  825&  48.6&  77.8\\ 
\bottomrule 
\end{tabular} 

	%}
\end{table}

Table~\ref{tab:bb} presents the results of the black-box testing on the 31 APIs described in Table~\ref{tab:gqlbapis}. 
The results include the number of endpoints (\#Endpoints) representing the number of queries and mutations present in the schema, the percentage of endpoints with generated tests without errors (\%NoErrors), and the percentage of endpoints with generated tests with errors (\%WithErrors). 
Tests with errors and others without errors could be generated for the same endpoint. 
However, the following formula would be satisfied: 

\begin{equation}
\frac{\%NoErrors}{100} + \frac{\%WithErrors}{100} \leq  2
\end{equation}
with
\begin{equation}
\%NoErrors = \frac{\#NoErrors}{\#Endpoints}
\end{equation}
and
\begin{equation}
\%WithErrors = \frac{\#WithErrors}{\#Endpoints}
\end{equation}

From Table~\ref{tab:bb}, we remark that all endpoints were reached, and responses with either \texttt{data} or \texttt{errors} fields are effectively derived. 
From the table we can see that we can generate tests which lead to responses with \texttt{errors} fields for many endpoints (77.8\%).
However, there are also many queries/mutations for which we could not get back any valid data (i.e., responses with \texttt{data} field and no \texttt{errors} were less than 50\%).
This is likely due to input constraints which are unlikely to be satisfied with random data.
Without code analysis (or constraints expressed directly on schema), likely there is not much a black-box tool can do here (besides having the user providing some sets of valid inputs to fuzz).  

Similarly to the fuzzing of RESTful APIs, black-box testing can find faults in GraphQL APIs by just sending random (but syntactically valid) inputs, as often APIs are not particularly robust when dealing with such kind of random data~\cite{marculescu2022faults}. 
However, without being able to analyze the source code, it can be hard to bypass their first layer of input validation and generate successful API requests~\cite{arcuri2020blackbox}.

\begin{result}
	\textbf{RQ2}: The Black-box testing implemented in our novel approach enables the automated test generation that can detect on average up to 641 endpoints with errors out of 825 endpoints (i.e., 77.8\%).  
\end{result}

%-------------------------------------------------------------------------------------------------
\subsubsection{Results for RQ3}

As discussed in Section~\ref{sub:GraphQl}, currently GraphQL makes no distinction between user and server errors. 
So, without an in-depth manual analysis of the generated tests, it is hard to tell which responses with \texttt{errors} messages are  due to actual software faults, and not a simple misuses of the API.
Furthermore, without knowing the full details of the expected business logic of the specific API under analysis, it might be hard for researchers (which are not the developers of the API) to determine if a returned error is indeed due to a software fault.
This problem is further exacerbated for the external APIs used for black-box testing experiments, where the source code is not available and cannot be used to validate if an error is indeed likely due to a fault.
Still, when evaluating a novel fuzzing technique like we do in this paper, it is important to check if it can find any actual faults.
For this reason, we did a manual analysis of hundreds of generated tests from our experiments.
Here, we discuss some of the most interesting cases.

Let us start from the following generated test case for \textit{petclinic}. 

\begin{lstlisting}[language=Java]
@Test(timeout = 60000)
public void test_4() throws Exception {
        
   given().accept("application/json")
          .contentType("application/json")
          .body(" { " + 
                    " \"query\": \"mutation{removeSpecialty(input:{specialtyId:643}){specialties{id}}}\" " + 
                    " } ")
          .post(baseUrlOfSut + "/graphql")
          .then()
          .statusCode(200) // org/springframework/samples/petclinic/repository/springdatajpa/SpringDataSpecialtyRepositoryImpl_38_delete
          .assertThat()
          .contentType("application/json")
          .body("'data'", nullValue())
          .body("'errors'.size()", equalTo(1))
          .body("'errors'[0].'message'", containsString("Internal Server Error(s) while executing query"))
          .body("'errors'[0].'path'", nullValue())
          .body("'errors'[0].'extensions'", nullValue());
}
\end{lstlisting}

Here,  the message \emph{``Internal Server Error(s) while executing query''} is a clear example of a fault, even if the returned HTTP status code is 200.
By debugging this test case, we found that the problem is due to a null pointer exception: trying to remove a specialty with id equals to 643 that does not exist.
Ideally, the API should return an error message stating the requested resource does not exist.
However, it looks like the implementation of such API is ignoring the cases when a user asks for something not in the database, which leads to an internal crash.

A similar case can be seen in the following test generated for \emph{react-finland}. 

\begin{lstlisting}[language=Java]
@Test(timeout = 60000)
public void test_0_with500() throws Exception {
        
   given().accept("application/json")
          .contentType("application/json")
          .body(" { " + 
                " \"query\": \"{theme(conferenceId:\\\"coM_FEt0ANyU87\\\"){id,fonts{primary}}} \" " + 
                " } ")
          .post(baseUrlOfSut)
          .then()
          .statusCode(500)
          .assertThat()
          .contentType("application/json")
          .body("'errors'.size()", equalTo(1))
          .body("'errors'[0].'message'", containsString("Conference id did not match series"))
          .body("'errors'[0].'locations'.size()", equalTo(1))
          .body("'errors'[0].'locations'[0].'line'", numberMatches(1.0))
          .body("'errors'[0].'locations'[0].'column'", numberMatches(5.0))
          .body("'errors'[0].'path'.size()", equalTo(1))
          .body("'errors'[0].'path'", hasItems("theme"))
          .body("'data'", nullValue());
}
\end{lstlisting}

Here, the API returns a meaningful error message stating \emph{``Conference id did not match series''}.
This means that the requested id \emph{``coM\_FEt0ANyU87''} is not matching any registered conference in the API.
However, the API is returning the HTTP status code 500, which in HTTP represents a server error, and not a user error (where the most suited code for this case would likely be 404).
Technically, this is a fault, although likely not a serious one.  
Not properly handling the requests for missing data seems common as well in RESTful APIs~\cite{marculescu2022faults}.

However, there are a few cases of more serious faults, like when the returned responses are not matching the constraints of the GraphQL schema of the API.
For example, consider the following case  of a HTTP call in a generated test for \emph{Bahnql}.

\begin{lstlisting}[language=Java]
given().accept("application/json")
       .contentType("application/json")
       .body(" { " + 
             " \"query\": \"{parkingSpace(id : 842)  {name,label,responsibility,spaceType,location{latitude},url,operator,distance,facilityType,openingHoursEn,isSpecialProductDb,isOutOfService,occupancy{validData,timestamp,timeSegment},clearanceHeight,outOfService,isMonthSeason,tariffDiscount,tariffPaymentCustomerCards,tariffFreeParkingTimeEn,tariffPaymentOptionsEn,slogan}       } \" " + 
              " } ")
       .post(baseUrlOfSut)
       .then()
       .statusCode(200)
       .assertThat()
       .contentType("application/json")
       .body("'errors'.size()", equalTo(2))
       .body("'errors'[0].'message'", containsString("Cannot return null for non-nullable field Location.latitude."))
       .body("'errors'[0].'locations'.size()", equalTo(1))
       .body("'errors'[0].'locations'[0].'line'", numberMatches(1.0))
       .body("'errors'[0].'locations'[0].'column'", numberMatches(77.0))
       .body("'errors'[0].'path'.size()", equalTo(3))
       .body("'errors'[0].'path'", hasItems("parkingSpace", "location", "latitude"))
       .body("'data'.'parkingSpace'", nullValue());
\end{lstlisting}

Here a 200 status code is returned, which would imply a success from the point of view of HTTP. 
However, the error message is \emph{``Cannot return null for non-nullable field Location.latitude.''}. 
This looks like a case of internal server error,
where a test case is asking for a non-nullable field named \emph{latitude}, but the server tried to return a \emph{null} value. 
All types in GraphQL are nullable by default, and the \emph{null} value is a valid response. 
However, when looking into its schema definition, the field \emph{latitude} is defined as a \emph{non-null} scalar.
This is a clear example showing an actual fault in the SUT, where the API tries to return a response that violates the schema. 

\begin{lstlisting}[language=Java]
 "kind": "OBJECT",
 "name": "Location",
 "fields":[
            {
              "name": "latitude",
              "args": [],
              "type": {
                "kind": "NON_NULL",
                "name": null,
                "ofType": {
                  "kind": "SCALAR",
                  "name": "Float",
                  "ofType": null
                }
              },
              "isDeprecated": false,
              "deprecationReason": null
            },
\end{lstlisting}

Another interesting example of schema violation is the following test generated for the \emph{Catalysis-hub} API.

\begin{lstlisting}[language=Java]
@Test(timeout = 60000)
public void test_11() throws Exception {
        
   given().accept("application/json")
          .contentType("application/json")
          .body(" { " + 
                " \"query\": \"{information(name:\\\"fyS8oO8Upt9ceuK\\\",value : \\\"CblJF7FCM_kT_\\\",distinct : false,op : \\\"4iyt\\\",search : \\\"Uk4WPkx7y\\\",jsonkey : \\\"dWws4\\\",order : \\\"pc\\\",before : \\\"\\\",after : \\\"CU1SdJQYVHfnce\\\",first : 298,last : 303)  {edges{node{value,id},cursor},totalCount}       } \" " + 
                " } ")
          .post(baseUrlOfSut)
          .then()
          .statusCode(200)
          .assertThat()
          .contentType("application/json")
          .body("'errors'.size()", equalTo(1))
          .body("'errors'[0].'message'", containsString("Can't find property named \"cursor\" on mapped class Information->information in this Query."))
          .body("'errors'[0].'locations'.size()", equalTo(1))
          .body("'errors'[0].'locations'[0].'line'", numberMatches(1.0))
          .body("'errors'[0].'locations'[0].'column'", numberMatches(5.0))
          .body("'errors'[0].'path'.size()", equalTo(1))
          .body("'errors'[0].'path'", hasItems("information"))
          .body("'data'.'information'", nullValue());
\end{lstlisting}

Here a 200 status code is returned, which would imply a success from the point of view of HTTP.
However in the body of the response the following error message appears: \emph{``Can't find property named "cursor" on mapped class Information->information in this Query''}.
It states that there is no field named \emph{cursor} belonging to the root query \emph{information}.
We have extracted and analyzed the whole schema of the \emph{Catalysis-hub} API by sending an introspective query to its endpoint.
The schema reveals that the field \emph{information} is of type \emph{InformationCountableConnection}.
The type \emph{InformationCountableConnection} has the field named \emph{edges} (that we have asked for) of type \emph{InformationCountableEdge}.
This latter, as shown below, has two fields, namely \emph{node} and \emph{cursor}, which shows a clear fault in the SUT.
The internal implementation of the server does not respect the defined GraphQL schema.
  
\begin{lstlisting}[language=Java]
          "kind": "OBJECT",
          "name": "InformationCountableEdge",
          "fields": [
            {
              "name": "node",
              "args": [],
              "type": {
                "kind": "OBJECT",
                "name": "Information",
                "ofType": null
              },
              "isDeprecated": false,
              "deprecationReason": null
            },
            {
              "name": "cursor",
              "args": [],
              "type": {
                "kind": "NON_NULL",
                "name": null,
                "ofType": {
                  "kind": "SCALAR",
                  "name": "String",
                  "ofType": null
                }
              },
              "isDeprecated": false,
              "deprecationReason": null
            }
          ],
          "inputFields": null,
          "interfaces": [],
          "enumValues": null,
          "possibleTypes": null
        },
\end{lstlisting}

Although schema violations might not be always easy to identify, there are other cases in which faults are very clear.
For instance, consider the following test generated for the \emph{Buildkite} API.

\begin{lstlisting}[language=Java]
given().accept("application/json")
       .header("Authorization", "Bearer 992ae7ae4998a8a8faa7c762d74e3c20f2abe154")
       .contentType("application/json")
       .body(" { " + 
             " \"query\": \"mutation{jobTypeBlockUnblock(input:{clientMutationId:\\\"hVHw\\\", id:\\\"R2qgRAwGnOo\\\", fields:\\\"6nj19G\\\"}){clientMutationId}}\" " + 
             " } ")
       .post(baseUrlOfSut)
       .then()
       .statusCode(200)
       .assertThat()
       .contentType("application/json")
       .body("'type'", containsString("unknown_error"))
       .body("'errors'.size()", equalTo(1))
       .body("'errors'[0].'message'", containsString("An error occurred while executing your GraphQL query. Please contact hello@buildkite.com for help and provide this query in the email."));
 \end{lstlisting}

Here, the server returned the status code 200.
However, an internal server error is detected by showing this message in the body: \emph{``An error occurred while executing your GraphQL query. Please contact hello@buildkite.com for help and provide this query in the email''}. 
Unfortunately, faults found by these black-box experiments from the requested APIs cannot be further analyzed, as we do not have access to the source code of this remote service.
It seems like a case of crash due to requesting a resource that does not exist, but we cannot be sure.

Another clear example of a major problem can be seen here in the following test generated for
the \emph{Catalysis-hub} API. 
The  test requests the resource called \emph{hasPreviousPage}, but the actual call (done with the library \texttt{RestAssured}) throws an exception.

\begin{lstlisting}[language=Java]
try{
    given().accept("application/json")
           .contentType("application/json")
           .body(" { " + 
                 " \"query\": \"{information(name:\\\"8qpQyCBnwDOgP2\\\",value:\\\"fU5gDkH6\\\",distinct: false,op : \\\"J0ZA3ahBYJTLgMp\\\",search : \\\"59W\\\",jsonkey : \\\"TMJOz\\\",order : \\\"aHdTeItZ\\\",before : \\\"g7_U2sanJ_l0\\\",after : \\\"bEUm\\\",first : 141,last : 847){pageInfo{hasPreviousPage}}} \" " + 
                 " } ")
           .post(baseUrlOfSut);
} catch(Exception e){
}
\end{lstlisting}

After a manual investigation, we found that the SUT is returning a failure in an HTML page instead of a JSON object, meaning a significant crush in the SUT. 

\begin{lstlisting}[language=Java]
"<!DOCTYPE html>\n\t<html>\n\t  <head>\n\t\t<meta name=\"viewport\" content=\"width=device-width, initial-scale=1\">\n\t\t<meta charset=\"utf-8\">\n\t\t<title>Application Error</title>\n\t\t<style media=\"screen\">\n\t\t  html,body,iframe {\n\t\t\tmargin: 0;\n\t\t\tpadding: 0;\n\t\t  }\n\t\t  html,body {\n\t\t\theight: 100%;\n\t\t\toverflow: hidden;\n\t\t  }\n\t\t  iframe {\n\t\t\twidth: 100%;\n\t\t\theight: 100%;\n\t\t\tborder: 0;\n\t\t  }\n\t\t</style>\n\t  </head>\n\t  <body>\n\t\t<iframe src=\"//www.herokucdn.com/error-pages/application-error.html\"></iframe>\n\t  </body>\n\t</html>"
\end{lstlisting}

However, this might had been due to hardware issues, and not a software fault in the business logic of the API.
For example, it could had well been that, although we did successfully generate an introspective query on this API, after a few hundreds HTTP calls the cloud provider Heroku (mentioned in that error page) temporarily disabled the API due to bandwidth usage constraints.

The last error we are going to discuss is for the following test case for the \emph{ecommerce-server} NodeJS API.
Note that, in contrast to the previous examples, the generated test here is in Jest (used for JavaScript/TypeScript APIs) format instead of JUnit. 

\begin{lstlisting}[language=Java]
test("test_6", async () => {

    let token_foo = "Bearer ";
    await superagent
            .post(baseUrlOfSut + "/graphql")
            .set('Content-Type','application/json')
            .send(" { " +
                " \"query\": \"mutation{login(data:{email:\\\"foo@foo.com\\\",password:\\\"bar123\\\"}){token}}\" " +
                " } ")
            .then(res => {token_foo += res.body.data.login.token;},
                error => {console.log(error.response.body); throw Error("Auth failed.")});;


    const res_0 = await superagent
            .post(baseUrlOfSut + "/graphql").set('Accept', "application/json")
            .set("Authorization", token_foo) // foo-auth
            .set('Content-Type','application/json')
            .send(" { " +
                " \"query\": \"  { findStoreById  (id : \\\"Z\\\")  {id,products{categories{name},brand,images},employees{createdAt,username,email,bio,image},createdAt,bio,rate,city,state,number,sales}       } \" " +
                " } ")
            .ok(res => res.status);

    expect(res_0.status).toBe(200); // build/src/store/store.service.js_39_40
    expect(res_0.header["content-type"].startsWith("application/json")).toBe(true);
    expect(res_0.body.errors.length).toBe(1);
    expect(res_0.body.errors[0].message).toBe("invalid input syntax for integer: \"Z\"");
    expect(res_0.body.errors[0].locations.length).toBe(1);
    expect(res_0.body.errors[0].locations[0].line).toBe(1.0);
    expect(res_0.body.errors[0].locations[0].column).toBe(5.0);
    expect(res_0.body.errors[0].path.length).toBe(1);
    expect(res_0.body.errors[0].path[0]).toBe("findStoreById");
    expect(res_0.body.errors[0].extensions.code).toBe("INTERNAL_SERVER_ERROR");
    expect(res_0.body.errors[0].extensions.exception.query).toBe("SELECT \"Store\".\"id\" AS \"Store_id\", \"Store\".\"created_at\" AS \"Store_created_at\", \"Store\".\"updated_at\" AS \"Store_updated_at\", \"Store\".\"name\" AS \"Store_name\", \"Store\".\"bio\" AS \"Store_bio\", \"Store\".\"rate\" AS \"Store_rate\", \"Store\".\"slug\" AS \"Store_slug\", \"Store\".\"street\" AS \"Store_street\", \"Store\".\"city\" AS \"Store_city\", \"Store\".\"state\" AS \"Store_state\", \"Store\".\"country\" AS \"Store_country\", \"Store\".\"neighborhood\" AS \"Store_neighborhood\", \"Store\".\"number\" AS \"Store_number\", \"Store\".\"zipCode\" AS \"Store_zipCode\", \"Store\".\"sales\" AS \"Store_sales\", \"Store__employees\".\"id\" AS \"Store__employees_id\", \"Store__employees\".\"created_at\" AS \"Store__employees_created_at\", \"Store__employees\".\"updated_at\" AS \"Store__employees_updated_at\", \"Store__employees\".\"name\" AS \"Store__employees_name\", \"Store__employees\".\"username\" AS \"Store__employees_username\", \"Store__employees\".\"email\" AS \"Store__employees_email\", \"Store__employees\".\"bio\" AS \"Store__employees_bio\", \"Store__employees\".\"image\" AS \"Store__employees_image\", \"Store__employees\".\"role\" AS \"Store__employees_role\", \"Store__employees\".\"status\" AS \"Store__employees_status\", \"Store__employees\".\"password\" AS \"Store__employees_password\", \"Store__products\".\"id\" AS \"Store__products_id\", \"Store__products\".\"created_at\" AS \"Store__products_created_at\", \"Store__products\".\"updated_at\" AS \"Store__products_updated_at\", \"Store__products\".\"title\" AS \"Store__products_title\", \"Store__products\".\"description\" AS \"Store__products_description\", \"Store__products\".\"brand\" AS \"Store__products_brand\", \"Store__products\".\"sku\" AS \"Store__products_sku\", \"Store__products\".\"price\" AS \"Store__products_price\", \"Store__products\".\"thumbnail\" AS \"Store__products_thumbnail\", \"Store__products\".\"images\" AS \"Store__products_images\", \"Store__products\".\"reviews\" AS \"Store__products_reviews\", \"Store__products\".\"quantity\" AS \"Store__products_quantity\", \"Store__products\".\"dimension\" AS \"Store__products_dimension\", \"Store__products\".\"storeId\" AS \"Store__products_storeId\" FROM \"store\" \"Store\" LEFT JOIN \"store_employees_users\" \"Store_Store__employees\" ON \"Store_Store__employees\".\"storeId\"=\"Store\".\"id\" LEFT JOIN \"users\" \"Store__employees\" ON \"Store__employees\".\"id\"=\"Store_Store__employees\".\"usersId\"  LEFT JOIN \"product\" \"Store__products\" ON \"Store__products\".\"storeId\"=\"Store\".\"id\" WHERE \"Store\".\"id\" IN ($1)");
    expect(res_0.body.errors[0].extensions.exception.parameters.length).toBe(1);
    expect(res_0.body.errors[0].extensions.exception.parameters[0]).toBe("Z");
    expect(res_0.body.errors[0].extensions.exception.driverError.length).toBe(90.0);
    expect(res_0.body.errors[0].extensions.exception.driverError.name).toBe("error");
    expect(res_0.body.errors[0].extensions.exception.driverError.severity).toBe("ERROR");
    expect(res_0.body.errors[0].extensions.exception.driverError.code).toBe("22P02");
    expect(res_0.body.errors[0].extensions.exception.driverError.file).toBe("numutils.c");
    expect(res_0.body.errors[0].extensions.exception.driverError.line).toBe("62");
    expect(res_0.body.errors[0].extensions.exception.driverError.routine).toBe("pg_atoi");
    expect(res_0.body.errors[0].extensions.exception.length).toBe(90.0);
    expect(res_0.body.errors[0].extensions.exception.severity).toBe("ERROR");
    expect(res_0.body.errors[0].extensions.exception.code).toBe("22P02");
    expect(res_0.body.errors[0].extensions.exception.file).toBe("numutils.c");
    expect(res_0.body.errors[0].extensions.exception.line).toBe("62");
    expect(res_0.body.errors[0].extensions.exception.routine).toBe("pg_atoi");
    expect(res_0.body.errors[0].extensions.exception.stacktrace.length).toBe(9);
    expect(res_0.body.errors[0].extensions.exception.stacktrace[0]).toBe("QueryFailedError: invalid input syntax for integer: \"Z\"");
    expect(res_0.body.errors[0].extensions.exception.stacktrace[1]).toBe("    at QueryFailedError.TypeORMError [as constructor] (D:\\WORK\\EXPERIMENTS\\graphql-journal\\y100k_0_9\\ecommerce-server\\node_modules\\typeorm\\error\\TypeORMError.js:9:28)");
    expect(res_0.body.errors[0].extensions.exception.stacktrace[2]).toBe("    at new QueryFailedError (D:\\WORK\\EXPERIMENTS\\graphql-journal\\y100k_0_9\\ecommerce-server\\node_modules\\typeorm\\error\\QueryFailedError.js:13:28)");
    // Skipping assertions on the remaining 6 elements. This limit of 3 elements can be increased in the configurations
    expect(res_0.body.data.findStoreById).toBe(null);
});
\end{lstlisting}

The API here returns the error message \emph{invalid input syntax for integer: ``Z''}.
In the schema,  the \texttt{id} input for the query \texttt{findStoreById} is of type string (and so \evo did send an input string like \emph{``Z''}), but the API is expecting an integer.
Technically, this could be considered as a schema-related fault.
However, there might be good reasons for sending a numeric value as a string in JSON, as JSON numbers are considered as 64-bit double-float values.  
This is an issue if one rather needs to deal with 64-bit integers (e.g., for numeric ids in SQL databases).
The major issue here, though, is ``where'' the check is done. 
The GraphQL specification allows to provide extra information in the \texttt{errors} objects, under an optional field called \texttt{extensions}. 
What can be present in this field depends on the different GraphQL framework implementations.
In this particular case, the full stack-trace of an internal thrown exception is added to the response. 

First, this could be technically a security issue if the API was in production and not just run locally for testing.
Full stack-trace details are useful for debugging, but they expose internal details of the API that could be exploited by external attackers.
Second, the exception seems to happen in a SQL SELECT query, which is malformed.
The point here is that the id in the SQL database is of type numeric, and so a value like \emph{``Z''} is invalid.
However, the API does not check for such integer constraints as a first layer of input validation when a GraphQL query is executed, and rather fail afterwards.
This can be a serious problem if there are modifications to the internal state of API before the thrown exception, as the API might be left in a inconsistent state.

\begin{result}
	\textbf{RQ3}: Different kinds of faults were automatically detected with our novel techniques, including wrong handling of requests for missing data, and generated responses that do not match the API schemas.
\end{result}

%%%%%%%%%%%%%%%%%%%%%%%%%%%%%%%%%%%%%%%%%%%%%%%%%%%%%%%%%%%%%%%%%%%%%%%%%%%%%%%%%%%%%%%%%%%%%%%%%%%%%%%%%%%%%%%%%%%%
\section{Discussion and Future Directions}
\label{sec:Discussion}
This section discusses the main findings of the paper, followed by possible future work. 
The main findings of using the \evo for automated GraphQL APIs testing can be summarized as follows:

\begin{enumerate}

\item The first finding of this study consists on the difficulty of automatically identifying test cases of GraphQL APIs compared to the RESTful APIs. 
Indeed, the graph representation of the actions is more complex than the traditional representation of the RESTful APIs. 
This representation is rich and might be used in different domain applications, however, this needs a careful care of the automated test generation process.
Furthermore, whereas a RESTful API can have clear relations between resources based on hierarchical URIs, and that information can be successfully exploited by test generation tools~\cite{zhang2019resource}, this does not seem the case for GraphQL APIs (e.g., no easy heuristics to determine which resources on the graph each mutation operation might manipulate). 

\item The second finding of this study is that the \evo tool proved its applicability in handling other kinds of web service APIs, represented by GraphQL APIs. 
\evo was implemented and architectured  from the start to be able to be extended and adapted to other system test generation domains besides REST APIs~\cite{arcuri2018evomaster}. 
 The results obtained in this paper shows that \evo is a generic enough framework for evolutionary-based system test generation, at least for applications where the entry point is a TCP connection.
Being released as open-source~\cite{arcuri2021evomaster}, \evo can be further extended and used in other domains as well.    

\item To obtain better code coverage, white-box heuristics based on search-based techniques can help significantly. However, existing APIs can have many faults that can be easily detected by simply sending random inputs. This makes even simple approaches like black-box testing potentially useful for practitioners. 

\end{enumerate}

In order to improve the effectiveness of the automated GraphQL API testing, several directions may be investigated in the future:

\begin{enumerate}
\item \textbf{Test oracle problem.} Given a test case, whether the result of its execution is correct or not can be determined with an automated \emph{oracle}~\cite{barr2015oracle}.
Without an automated oracle, the developer has to determine manually whether the observed  test results are as expected or not. 
But having to manually check hundreds/thousands of generated test cases might not viable. 
As discussed in the paper, query responses with \texttt{errors} fields might not be representing actual faults in the SUT, but rather just the user sending wrong data.
In order to mitigate the test oracle problem, an intelligent automated strategy is needed to differentiate between the actual faults from the user errors for a given GraphQL API response. 
One approach is to use machine learning, in particular supervised classification, to automatically label whether a response with  \texttt{errors} field should be treated as a potential fault that the developer should investigate. 
When test suites are generated at the end of the search, the test cases could be ordered based on their probability of representing actual faults.

\item \textbf{Evolutionary Computation.}
Evolutionary computation is an intelligent mechanism of exploring large and big solution spaces, inspired by the evolutionary process from nature. 
In this research work, we only used the evolutionary algorithm MIO already present in \evo, but others might be more fitting for the case of GraphQL API testing.   
In order to further improve the code coverage of the automated testing in this domain, further investigation should be carried out in this area.
For example, other techniques such as Particle Swarm Optimization~\cite{hossain2016big} and Ant Colony Optimization~\cite{wu2013transactional} could be considered. 
Combining other testing techniques (e.g., Symbolic Execution~\cite{baldoni2018survey}) with  evolutionary computation can also be considered a good direction to further address this problem~\cite{galeotti2014extending}.

\item \textbf{Knowledge Discovery.} Data mining and knowledge discovery is the process of extracting hidden patterns from a large data collection. 
Decomposition is a widely used technique in solving complex problems~\cite{djenouri2021exploring,djenouri2018fast}. 
The aim is to create highly correlated clusters, where each cluster contains similar data. 
In our context, the idea is to apply the decomposition method to the GraphQL schema in order to derive sub-graphs of schema. 
Each sub-graph might contain highly connected actions. 
Good decomposition methods allow to find independent sub-graphs as much as possible, in order to enable the same test case generation while dealing with the sub-graphs as when dealing with the entire GraphQL schema.

\item \textbf{Industrial Settings.} Further investigations with more case studies  will be essential to generalize the effectiveness of our novel technique. 
Of particular importance it will be to apply our technique in industrial settings, to see and evaluate how engineers would use tools like \evo in practice on their APIs.
\end{enumerate}

%%%%%%%%%%%%%%%%%%%%%%%%%%%%%%%%%%%%%%%%%%%%%%%%%%%%%%%%%%%%%%%%%%%%%%%%%%%%
\section{Threats To Validity}
\label{sec:threats}

Threats to internal validity come from the fact that our experiments are derived from a software tool.
Errors in such a tool could negatively affect the validity of our empirical results.
Although our \evo extension was carefully tested, we cannot provide any guarantee of not having software faults.
However, as it is open-source, anyone can review its source code. 
Another potential issue is that the implemented solution in this research work is based on random algorithms.
This happens in particular for population initialization of the evolutionary algorithm, where different test cases may be generated.
To deal with this issue, each experiment for white-box testing was repeated 30 times~\cite{Hitchhiker14}, with different random seeds, and the appropriate statistical tests were used to analyze the results.
All the APIs used for the white-box experiments are collected in a GitHub repository called EMB~\cite{EMB}, which is stored on Zenodo as well~\cite{andrea_arcuri_2022_6106830}.
Furthermore, all of our scripts used to carry out our experiments are stored as part of the repository of \evo.
This is done to enable third-parties to replicate and validate our experiments.
However, experiments for black-box testing cannot be reliably replicated, as they rely on live services on which we do not have any control on (e.g., they can be modified at any time by their owners).

Threats to external validity are due to the fact that only 7 GraphQL APIs for white-box testing, and 31 GraphQL APIs for black-box testing, were used in our empirical analysis. 
The generalization of such results to other APIs might not be possible at this stage.
More APIs should be investigated in the future.
However, as this is the first work on white-box testing of GraphQL APIs, already achieving good coverage and finding real faults on a complex GraphQL API provide a promising first step.

%%%%%%%%%%%%%%%%%%%%%%%%%%%%%%%%%%%%%%%%%%%%%%%%%%%%%%%%%%%%%%%%%%%%%%%%%%%%
\section{Conclusions}
\label{sec:conclusions}

This paper introduced a new approach for automated testing for GraphQL APIs. 
It is a full complete solution, starting from the schema extraction and ending by automatically generating test cases outputted in JUnit and Jest format. Two testing modes are implemented and evaluated: white-box and black box testing.    

In order to intelligently explore the test case space, evolutionary computation techniques are used in the white-box testing.
Two mutation operators (internal and structure mutation) are defined, where the goal is to maximize code coverage and fault-finding. In addition, random testing is used for the black-box mode.  

To validate the applicability of the proposed framework, it is integrated into the \evo open-source tool. 
Our empirical analysis was carried out on 7 GraphQL APIs for white-box testing, and 31 GraphQL APIs for black-box testing. 
The results show the clear improvement of using the evolutionary computation compared with the random search baseline for white-box testing.
Regarding black-box testing, several real faults were found by random testing in the analyzed APIs. 

To learn more about \evo, visit \texttt{www.evomaster.org}.

%%%%%%%%%%%%%%%%%%%%%%%%%%%%%%%%%%%%%%%%%%%%%%%%%%%%%%%%%%%%%%%%%%%%%%%%%%%%
\section*{Acknowledgments}
This work is funded by the European Research Council (ERC) under the European Union’s Horizon 2020 research and innovation programme (EAST project, grant agreement No. 864972).

%%%%%%%%%%%%%%%%%%%%%%%%%%%%%%%%%%%%%%%%%%%%%%%%%%%%%%%%%%%%%%%%%%%%%%%%%%%%
%https://arxiv.org/help/submit_tex#latex
% For arXiv, use generated bbl
%\bibliographystyle{acm}
%\bibliography{papers}

\begin{thebibliography}{10}

\bibitem{gqlapisguru}
apis.guru.
\newblock https://apis.guru/graphql-apis/.

\bibitem{gqlecommerce}
e-commerce.
\newblock https://github.com/react-shop/react-ecommerce.

\bibitem{EvoMaster}
{EvoMaster}.
\newblock https://github.com/EMResearch/EvoMaster.

\bibitem{EMB}
Evomaster benchmark (emb).
\newblock https://github.com/EMResearch/EMB.

\bibitem{Github}
Github.
\newblock https://github.com.

\bibitem{GraphQLFoundation}
Graphql foundation.
\newblock https://graphql.org/foundation/.

\bibitem{gqlpatio}
Patio-api.
\newblock https://github.com/patio-team/patio-api.

\bibitem{gqlpetclinic}
petclinic.
\newblock https://github.com/spring-petclinic/spring-petclinic-graphql.

\bibitem{gqlreact-finland}
react-finland.
\newblock https://github.com/ReactFinland/graphql-api.

\bibitem{gqltimbuctoo}
timbuctoo.
\newblock https://github.com/HuygensING/timbuctoo.

\bibitem{ali2009systematic}
{\sc Ali, S., Briand, L.~C., Hemmati, H., and Panesar-Walawege, R.~K.}
\newblock A systematic review of the application and empirical investigation of
  search-based test case generation.
\newblock {\em IEEE Transactions on Software Engineering 36}, 6 (2009),
  742--762.

\bibitem{Alshraideh06}
{\sc Alshraideh, M., and Bottaci, L.}
\newblock Search-based software test data generation for string data using
  program-specific search operators.
\newblock {\em Software Testing, Verification, and Reliability 16}, 3 (2006),
  175--203.

\bibitem{mio2017}
{\sc Arcuri, A.}
\newblock {Many Independent Objective (MIO) Algorithm for Test Suite
  Generation}.
\newblock In {\em International Symposium on Search Based Software Engineering
  (SSBSE)\/} (2017), pp.~3--17.

\bibitem{arcuri2018evomaster}
{\sc Arcuri, A.}
\newblock {EvoMaster: Evolutionary Multi-context Automated System Test
  Generation}.
\newblock In {\em IEEE International Conference on Software Testing,
  Verification and Validation (ICST)\/} (2018), IEEE.

\bibitem{arcuri2018experience}
{\sc Arcuri, A.}
\newblock An experience report on applying software testing academic results in
  industry: we need usable automated test generation.
\newblock {\em Empirical Software Engineering 23}, 4 (2018), 1959--1981.

\bibitem{arcuri2019restful}
{\sc Arcuri, A.}
\newblock Restful api automated test case generation with evomaster.
\newblock {\em ACM Transactions on Software Engineering and Methodology (TOSEM)
  28}, 1 (2019), 3.

\bibitem{arcuri2020blackbox}
{\sc Arcuri, A.}
\newblock Automated black-and white-box testing of restful apis with evomaster.
\newblock {\em IEEE Software 38}, 3 (2020), 72--78.

\bibitem{ArB11a}
{\sc Arcuri, A., and Briand, L.}
\newblock Adaptive random testing: An illusion of effectiveness?
\newblock In {\em ACM Int. Symposium on Software Testing and Analysis
  (ISSTA)\/} (2011), pp.~265--275.

\bibitem{Hitchhiker14}
{\sc Arcuri, A., and Briand, L.}
\newblock {A Hitchhiker's Guide to Statistical Tests for Assessing Randomized
  Algorithms in Software Engineering}.
\newblock {\em Software Testing, Verification and Reliability (STVR) 24}, 3
  (2014), 219--250.

\bibitem{arcuri2020sql}
{\sc Arcuri, A., and Galeotti, J.~P.}
\newblock Handling sql databases in automated system test generation.
\newblock {\em ACM Transactions on Software Engineering and Methodology (TOSEM)
  29}, 4 (2020), 1--31.

\bibitem{arcuri2020testability}
{\sc Arcuri, A., and Galeotti, J.~P.}
\newblock Testability transformations for existing apis.
\newblock In {\em 2020 IEEE 13th International Conference on Software Testing,
  Validation and Verification (ICST)\/} (2020), IEEE, pp.~153--163.

\bibitem{arcuri2021evomaster}
{\sc Arcuri, A., Galeotti, J.~P., Marculescu, B., and Zhang, M.}
\newblock Evomaster: A search-based system test generation tool.
\newblock {\em Journal of Open Source Software 6}, 57 (2021), 2153.

\bibitem{AIB11}
{\sc Arcuri, A., Iqbal, M.~Z., and Briand, L.}
\newblock Random testing: Theoretical results and practical implications.
\newblock {\em IEEE Transactions on Software Engineering (TSE) 38}, 2 (2012),
  258--277.

\bibitem{andrea_arcuri_2022_6651631}
{\sc Arcuri, A., ZhangMan, asmab89, Bogdan, Gol, A., Galeotti, J.~P., Seran,
  López, A.~M., Aldasoro, A., Panichella, A., and Niemeyer, K.}
\newblock Emresearch/evomaster:, June 2022.

\bibitem{andrea_arcuri_2022_6106830}
{\sc Arcuri, A., ZhangMan, Gol, A., and asmab89}.
\newblock Emresearch/emb:, Feb. 2022.

\bibitem{baldoni2018survey}
{\sc Baldoni, R., Coppa, E., D’elia, D.~C., Demetrescu, C., and Finocchi, I.}
\newblock A survey of symbolic execution techniques.
\newblock {\em ACM Computing Surveys (CSUR) 51}, 3 (2018), 1--39.

\bibitem{barr2015oracle}
{\sc Barr, E.~T., Harman, M., McMinn, P., Shahbaz, M., and Yoo, S.}
\newblock The oracle problem in software testing: A survey.
\newblock {\em IEEE Transactions on Software Engineering (TSE) 41}, 5 (2015),
  507--525.

\bibitem{belhadi2022graphql}
{\sc Belhadi, A., Zhang, M., and Arcuri, A.}
\newblock {Evolutionary-based Automated Testing for GraphQL APIs}.
\newblock In {\em Genetic and Evolutionary Computation Conference (GECCO)\/}
  (2022).

\bibitem{cabrera2020towards}
{\sc Cabrera, E., C{\'a}rdenas, P., Cedillo, P., and Pes{\'a}ntez-Cabrera, P.}
\newblock Towards a methodology for creating internet of things (iot)
  applications based on microservices.
\newblock In {\em 2020 IEEE International Conference on Services Computing
  (SCC)\/} (2020), IEEE, pp.~472--474.

\bibitem{cirillo2020smart}
{\sc Cirillo, F., G{\'o}mez, D., Diez, L., Maestro, I.~E., Gilbert, T. B.~J.,
  and Akhavan, R.}
\newblock Smart city iot services creation through large-scale collaboration.
\newblock {\em IEEE Internet of Things Journal 7}, 6 (2020), 5267--5275.

\bibitem{djenouri2018fast}
{\sc Djenouri, Y., Belhadi, A., Fournier-Viger, P., and Lin, J. C.-W.}
\newblock Fast and effective cluster-based information retrieval using frequent
  closed itemsets.
\newblock {\em Information Sciences 453\/} (2018), 154--167.

\bibitem{djenouri2021exploring}
{\sc Djenouri, Y., Lin, J. C.-W., N{\o}rv{\aa}g, K., Ramampiaro, H., and Yu,
  P.~S.}
\newblock Exploring decomposition for solving pattern mining problems.
\newblock {\em ACM Transactions on Management Information Systems (TMIS) 12}, 2
  (2021), 1--36.

\bibitem{fraser2011evosuite}
{\sc Fraser, G., and Arcuri, A.}
\newblock Evo{S}uite: automatic test suite generation for object-oriented
  software.
\newblock In {\em ACM Symposium on the Foundations of Software Engineering
  (FSE)\/} (2011), pp.~416--419.

\bibitem{galeotti2014extending}
{\sc Galeotti, J.~P., Fraser, G., and Arcuri, A.}
\newblock Extending a search-based test generator with adaptive dynamic
  symbolic execution.
\newblock In {\em ACM Int. Symposium on Software Testing and Analysis
  (ISSTA)\/} (2014), ACM, pp.~421--424.

\bibitem{godefroid2020differential}
{\sc Godefroid, P., Lehmann, D., and Polishchuk, M.}
\newblock Differential regression testing for rest apis.
\newblock In {\em Proceedings of the 29th ACM SIGSOFT International Symposium
  on Software Testing and Analysis\/} (2020), pp.~312--323.

\bibitem{harman2012search}
{\sc Harman, M., Mansouri, S.~A., and Zhang, Y.}
\newblock Search-based software engineering: Trends, techniques and
  applications.
\newblock {\em ACM Computing Surveys (CSUR) 45}, 1 (2012), 11.

\bibitem{hartig2018semantics}
{\sc Hartig, O., and P{\'e}rez, J.}
\newblock Semantics and complexity of graphql.
\newblock In {\em Proceedings of the 2018 World Wide Web Conference\/} (2018),
  pp.~1155--1164.

\bibitem{hossain2016big}
{\sc Hossain, M.~S., Moniruzzaman, M., Muhammad, G., Ghoneim, A., and Alamri,
  A.}
\newblock Big data-driven service composition using parallel clustered particle
  swarm optimization in mobile environment.
\newblock {\em IEEE Transactions on Services Computing 9}, 5 (2016), 806--817.

\bibitem{karlsson2020automatic}
{\sc Karlsson, S., {\v{C}}au{\v{s}}evi{\'c}, A., and Sundmark, D.}
\newblock Automatic property-based testing of graphql apis.
\newblock {\em arXiv preprint arXiv:2012.07380\/} (2020).

\bibitem{kim2022arxiv}
{\sc Kim, M., Xin, Q., Sinha, S., and Orso, A.}
\newblock Automated test generation for rest apis: No time to rest yet, 2022.

\bibitem{Kim2022Rest}
{\sc Kim, M., Xin, Q., Sinha, S., and Orso, A.}
\newblock Automated test generation for rest apis: No time to rest yet.
\newblock In {\em Proceedings of the 31st ACM SIGSOFT International Symposium
  on Software Testing and Analysis\/} (New York, NY, USA, 2022), ISSTA 2022,
  Association for Computing Machinery, p.~289–301.

\bibitem{Kor90}
{\sc Korel, B.}
\newblock Automated software test data generation.
\newblock {\em IEEE Transactions on software engineering 16}, 8 (1990),
  870--879.

\bibitem{mao2016sapienz}
{\sc Mao, K., Harman, M., and Jia, Y.}
\newblock {Sapienz: Multi-objective automated testing for android
  applications}.
\newblock In {\em ACM Int. Symposium on Software Testing and Analysis
  (ISSTA)\/} (2016), ACM, pp.~94--105.

\bibitem{marculescu2022faults}
{\sc Marculescu, B., Zhang, M., and Arcuri, A.}
\newblock On the faults found in rest apis by automated test generation.
\newblock {\em ACM Transactions on Software Engineering and Methodology (TOSEM)
  31}, 3 (2022), 1--43.

\bibitem{martin2021specification}
{\sc Martin-Lopez, A., Segura, S., Muller, C., and Ruiz-Cort{\'e}s, A.}
\newblock Specification and automated analysis of inter-parameter dependencies
  in web apis.
\newblock {\em IEEE Transactions on Services Computing\/} (2021).

\bibitem{newman2015building}
{\sc Newman, S.}
\newblock {\em Building Microservices}.
\newblock " O'Reilly Media, Inc.", 2015.

\bibitem{taelman2018graphql}
{\sc Taelman, R., Vander~Sande, M., and Verborgh, R.}
\newblock Graphql-ld: linked data querying with graphql.
\newblock In {\em ISWC2018, the 17th International Semantic Web Conference\/}
  (2018), pp.~1--4.

\bibitem{vargas2018deviation}
{\sc Vargas, D.~M., Blanco, A.~F., Vidaurre, A.~C., Alcocer, J. P.~S., Torres,
  M.~M., Bergel, A., and Ducasse, S.}
\newblock Deviation testing: A test case generation technique for graphql apis.
\newblock In {\em 11th International Workshop on Smalltalk Technologies
  (IWST)\/} (2018), pp.~1--9.

\bibitem{viglianisi2020resttestgen}
{\sc Viglianisi, E., Dallago, M., and Ceccato, M.}
\newblock Resttestgen: Automated black-box testing of restful apis.
\newblock In {\em IEEE International Conference on Software Testing,
  Verification and Validation (ICST)\/} (2020), IEEE.

\bibitem{wu2013transactional}
{\sc Wu, Q., and Zhu, Q.}
\newblock Transactional and qos-aware dynamic service composition based on ant
  colony optimization.
\newblock {\em Future Generation Computer Systems 29}, 5 (2013), 1112--1119.

\bibitem{zetterlund2022harvesting}
{\sc Zetterlund, L., Tiwari, D., Monperrus, M., and Baudry, B.}
\newblock Harvesting production graphql queries to detect schema faults.
\newblock In {\em 2022 IEEE Conference on Software Testing, Verification and
  Validation (ICST)\/} (2022), IEEE, pp.~365--376.

\bibitem{zhang2021adaptive}
{\sc Zhang, M., and Arcuri, A.}
\newblock Adaptive hypermutation for search-based system test generation: A
  study on rest apis with evomaster.
\newblock {\em ACM Transactions on Software Engineering and Methodology (TOSEM)
  31}, 1 (2021).

\bibitem{zhang2022arxiv}
{\sc Zhang, M., and Arcuri, A.}
\newblock Open problems in fuzzing restful apis: A comparison of tools, 2022.

\bibitem{zhang2022open}
{\sc Zhang, M., and Arcuri, A.}
\newblock Open problems in fuzzing restful apis: A comparison of tools.
\newblock {\em arXiv preprint arXiv:2205.05325\/} (2022).

\bibitem{meituanArxiv2022}
{\sc Zhang, M., Arcuri, A., Li, Y., Xue, K., Wang, Z., Huo, J., and Huang, W.}
\newblock Fuzzing microservices in industry: Experience of applying evomaster
  at meituan, 2022.

\bibitem{js2022}
{\sc Zhang, M., Belhadi, A., and Arcuri, A.}
\newblock Javascript instrumentation for search-based software testing: A study
  with restful apis.
\newblock In {\em IEEE International Conference on Software Testing,
  Verification and Validation (ICST)\/} (2022), IEEE.

\bibitem{zhang2019resource}
{\sc Zhang, M., Marculescu, B., and Arcuri, A.}
\newblock Resource-based test case generation for restful web services.
\newblock In {\em Proceedings of the Genetic and Evolutionary Computation
  Conference\/} (2019), pp.~1426--1434.

\end{thebibliography}

\end{document}